# Graphene synthesis and band gap opening

Deep Jariwala[a,+], Anchal Srivastava[b,*] and Pulickel M. Ajayan[c]

Graphene- the wonder material has attracted a great deal of attention from varied fields of condensed matter physics, materials science and chemistry in recent times. Its 2D atomic layer structure and unique electronic band structure makes it attractive for many applications. Its high carrier mobility, high electrical and thermal conductivity make it an exciting material. However, its applicability cannot be effectively realised unless facile techniques to synthesize high quality, large area graphene are developed in a cost effective way. Besides that a great deal of effort is required to develop techniques for modifying and opening its band structure so as to make it a potential replacement for silicon in future electronics. Considerable research has been carried out for synthesizing graphene and related materials by a variety of processes and at the same time a great deal of work has also taken place for manipulating and opening its electronic band structure. This review summarizes recent developments in the synthesis methods for graphene. It also summarizes the developments in graphene nanoribbon synthesis and methods to open band gap in graphene, in addition to pointing out a direction for future research and developments.

[a]    Department of Metallurgical Engineering, Institute of Technology,

       Banaras Hindu University, Varanasi-221005, India.

[+] Present Address: Department of Materials Science and Engineering,

     Northwestern University, Evanston, Illinois-60208, USA

 [b]    Department of Physics, Faculty of Science,

       Banaras Hindu University, Varanasi-221005, India

       E-mail: anchalbhu@gmail.com

       * Corresponding author

 [c]    Mechanical Engineering and Materials Science department, Rice

       University, Houston, Texas-77005, USA.

# Table of contents:

# 1.Introduction

## 1.1 Allotropes of carbon and their description:

Carbon is the 15th most abundant element in the Earth's crust, and the fourth most abundant element in the universe by mass after hydrogen, helium, and oxygen. It is present in all known life forms, and in the human body carbon is the second most abundant element by mass (about 18.5%) after oxygen. This abundance, together with the unique diversity of organic compounds and their unusual polymer-forming ability at temperatures commonly encountered on Earth, make this element the chemical basis of all known life. There are several solid state allotropes of carbon of which the best known are graphite and diamond; mixtures of bonding of carbon found in graphite and diamond give rise to many other forms such as amorphous carbon, diamond-like carbon, fullerenes, carbon nanotubes etc. The amorphous form is an assortment of carbon atoms in a non-crystalline, irregular, glassy state with essentially graphitic bonding with no long range order. It is present as a powder, and is the main constituent of substances such as charcoal, lampblack (soot) and activated carbon. At very high pressures carbon forms the more compact allotrope diamond, having nearly twice the density of graphite. Diamond has the same cubic structure as silicon and germanium. The carbon is bonded tetrahedrally ($sp^3$ bonding) in diamond making it isotropic. Diamond is in fact thermodynamically unstable under normal conditions and transforms into graphite. But due to a high activation energy barrier, the transition into graphite is extremely slow at room temperature as to be unnoticeable. Graphite holds the distinction of being the most stable form of carbon under standard conditions. Graphite has a layered structure. In each layer, the carbon atoms are arranged in a hexagonal lattice with separation of 0.142 nm, and the distance between planes is 0.335 nm. The in plane bonding is covalent between carbon atoms ($sp^2$ hybridization) and is therefore very strong. However inter planar bonding in graphite is mainly due to Van der Waals forces and hence very weak. A single layer of hexagonally arranged carbon atoms in graphite is called graphene. Fullerenes and carbon nanotubes are derivatives of graphite (or graphene). Thus graphene is the mother of all graphitic carbon forms.

## 1.2 History of graphene research and its applications

Studies on few layer graphite had begun long ago in late 1970's[1-2]. These studies were primarily based on epitaxial growth of few layer graphite films on transition metal surfaces by decomposing organic gases. More progress was made in the study of thin epitaxial films on catalytic metal surfaces throughout the 90's with the help of advanced surface characterization techniques like LEED, AES, HREELS etc. These studies have been exhaustively described in a review article by Oshima and Nagashima[3]. However no studies on electronic properties of these ultrathin graphite films were reported since these were all synthesized on conductive metallic substrates. This area received a sudden jolt in 2004 when Geim and co-workers at the University of Manchester were successful in isolating single layer graphite on an insulating substrate ($SiO_2$)[4]. This enabled the first report on electrical properties of single layer graphite or graphene and this started a whole new era in graphene research. The wealth of new scientific knowledge obtained by studying this system was so much that Geim and Novoselov were jointly awarded the Nobel Prize in Physics, 2010 within just 7 years of their first paper in the field[4]. This goes on to show the amount of promise this material carries both from fundamental science as well as technological and applications perspective. More interest was generated in graphene due to experimental observation of exciting properties. A very high intrinsic strength and Young's modulus of ~1100 GPa has been reported for graphene[5]. Researchers have also estimated a very high thermal conductivity of ~5000 Watt $m^{-1}$ $K^{-1}$ for monolayer graphene[6]. Single layer graphene absorbs 2.3% of incident light which is very high considering only one layer thickness[7]. ultra high electron mobility[8], very high specific surface area[9] and room temperature Quantum Hall Effect[10] have also been reported on single layer graphene.

Due to the above mentioned properties graphene finds a variety of applications. Most importantly in field effect transistors its very high mobility ( 10,000-1,40,000 $cm^2\ V^{-1}s^{-1}$ )[11][12] is attractive although is also a zero band gap semiconductor which means that inducing a bandgap is intrinsic to graphene based transistors. Graphene based composite materials show high electrical conductivity[13] and have found potential applications in biosensing[14] as well as Li ion batteries[15]. Due to very low resistance and high electron mobility graphene nanoribbons also have the potential to replace copper as interconnects in microelectronics [16]. Chemically derived graphene has been used to fabricate ultracapacitors with very high energy density[9]. Graphene based sensors are also very promising because a sharp dependence of electrical conductivity of graphene on concentration of adsorbates. Transparent conducting graphene films can be used as transparent electrodes in photovoltaic applications[17-18], Liquid Crystal Displays (LCDs)[19] as well as Organic Light Emitting Diodes (OLEDs). For such applications graphene films can potentially replace ITO glass films since graphene's mechanical strength and flexibility are superior compared to Indium Tin Oxide (ITO)[20-21].

**1.3 Focus of this review**

The commercial synthesis of graphene and graphene based products is still in its infancy. Moreover due to its zero band gap electronic structure, the use of graphene gets severely limited in FETs as it gives a low switching ratio. Thus opening up of band gap in graphene remains another challenge. This review article will focus on recent developments in the area of large scale, cheap and facile synthesis of high quality graphene which can be used for electronics as well as developments in various techniques used for doping and opening a finite band gap in graphene.

**About the authors:**

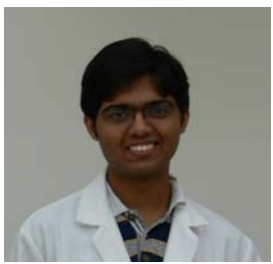

Deep Jariwala completed his B.Tech in Metallurgical Engineering from Institute of Technology, Banaras Hindu University in 2010. He has worked in Prof Ajayan's group at Rice University as a summer intern in 2008 and 2009 where he researched on graphene nanocutting, CVD synthesis and doping. He is currently pursuing his PhD in materials science and engineering at Northwestern Univeristy.

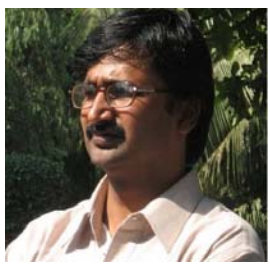

Anchal Srivastava received his Ph.D from Banaras Hindu University in 2002. He continued in the same department as a post doctoral fellow and later joined as a lecturer in 2004. He is currently serving as an assistant professor at Dept. of Physics, BHU. He specializes in the area of synthesis, characterizations and applications of carbon nanomaterials. He has been awarded the Max Planck India fellowship and carried out his research at Max Planck Institute of Solid State Physics, Stuttgart, Germany in Prof. Klaus Von Klitzing's group. He has also been awarded the prestigious BOYSCAST fellowship from government of India in 2008-09 during which he carried out research on graphene and related materials at Prof. P. M. Ajayan's group in Rice University. He has also been a visiting research scholar at Rensselaer Polytechnic Institute, USA in 2006.

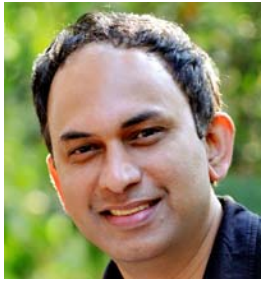

Professor Ajayan earned his B. Tech in Metallurgical engineering from Banaras Hindu University in 1985 and Ph.D. in materials science and engineering from Northwestern University in 1989. After three years of post-doctoral experience at NEC Corporation in Japan, he spent two years as a research scientist at the Laboratoire de Physique des Solides, Orsay in France and nearly a year and a half as an Alexander von Humboldt fellow at the Max-Planck-Institut fur Metallforschung, Stuttgart in Germany. In 1997, he joined the materials science and engineering faculty at Rensselaer as an Assistant Professor and was the Henri Burlage chair Professor in Engineering until he left RPI in 2007. He joined the mechanical engineering and materials science department of Rice University, as the Benjamin M. and Mary Greenwood Anderson Professor in Engineering from July 2007. Professor Ajayan's research interests include synthesis and structure-property relations of nanostructures and nanocomposites, materials science and applications of nanomaterials, energy storage, and phase stability in nanoscale systems. He is one of the pioneers in the field of carbon nanotubes and was involved in the early work on the topic along with the NEC group. He has published one book and 325 journal papers with more than 21,500 citations and an h-index of 75.

## 2. Synthesis of graphene

Various techniques have been found for producing thin graphitic films. It has been known since the late 1970's that carbon precipitates in the form of thin graphitic layers on transition metal surfaces[1-2]. However as mentioned earlier never were their electronic properties investigated because of the difficulty in isolating and transferring them onto insulating substrates. But in the late 90's Ruoff and co workers tried isolating thin graphitic flakes on SiO2 substrates by mechanical rubbing of patterned islands on HOPG[22]. However there was no report on their electrical property characterization. Using a similar method this was later achieved in 2005 by Kim and co workers and the electrical properties were reported[23]. However the real gold rush in graphene research began after Geim and co workers first published their work of isolating graphene onto $SiO_2$ substrate and measuring its electrical properties. After discovery of graphene in 2004 various techniques were developed to produce thin graphitic films and few layer graphene. They will be discussed in detail below.

**2.1 Epitaxial growth on metals**

Studies on ultrathin graphitic films epitaxially grown on metal surfaces have been carried out for a while[1, 24-33]. Same metals have been used to grow epitaxial single layer graphene films more recently [34-42]. Most of these metals show catalytic activity for hydrocarbon decomposition at high temperatures. The growth occurs by the principle of segregation of carbon[38], C vapor deposition[43] or by chemical vapor deposition which involves hydrocarbon decomposition on the catalytic metal surface[44]. In all these studies the growth was carried out on metallic single crystals under UHV conditions. For segregation controlled growth, the metal single crystal is raised to high temperatures under high vacuum where the bulk solubility of interstitial carbon in the metal lattice is high. After that the metal crystal is slowly cooled and the solubility sharply decreases thus segregating carbon at the surface which grows into graphene films. A schematic of the growth process is shown in Fig.(1) b. A detailed study of the mechanism, kinetics and dynamics of epitaxial graphene growth on metallic single crystals has been carried out by K F McCarty and co workers [43, 45-46]. According to their observations the rate of graphene nucleation and growth is strongly dependent on surface C adatom concentration while the limiting factor in graphene growth kinetics is C attachment to the graphene islands rather than surface diffusivity of C adatoms. Moreover they also propose that growth rate is favorable is orientation dependent. This highlights the fact that the arrangement of C atoms at the edges of the island and their attachment with the substrate plays a vital role in graphene growth. Another interesting observation is that the growth kinetics are non linear which suggests that a carbon atom cluster attachment to the graphene nucleus is more favorable than C monomer attachment. They also report that graphene growth occurs only by surface adatom diffusion and not by bulk diffusion and also that growth rate is independent of the C adatom source. A study on graphene growth dynamics by time resolved ARPES (Angle Resolved Photo Emission Spectroscopy) has also been reported[47]. Due to mismatch between the lattices of the metallic substrate and graphene a periodic moiré pattern was observed on the epitaxial graphene[3, 34, 36-37, 39, 44, 48]. These Moiré superstructures on epitaxially grown monolayer graphite have been reported as early as 1992[44]. However these moiré patterns are observed in STM investigations only when the lattice mismatch is greater than 1% as in the case of Ir, Ru, Pt and Pd whereas they are not observed in case of epitaxial growth on Ni[25, 35] and Co[35, 49]. It has also been observed that these periodic perturbations or Moire fringes due to lattice mismatch can be eliminated if graphene and the metal surface are intercalated with adsorbed oxygen on metal surface[50]. Epitaxial graphene films on metallic single crystals were mostly characterized insitu in UHV chambers. The most popular techniques for their characterization include Low Energy Electron Microscopy (LEEM), Reflected High Energy Electron Diffraction (RHEED), Auger electron spectroscopy (AES) and Low Energy Electron Diffraction (LEED). AFM and STM analysis was also carried out in some studies to study the Moire fringes. LEED and LEEM (or RHEED) are used together to obtain images and diffraction patterns simultaneously of the substrate thus giving one the ability to correlate a change in image with the graphene growth. AES is normally used to detect and map surface C adatom concentrations as function of time and temperature thus providing valuable data to understand nucleation and growth mechanisms. Since single crystal metallic substrates and UHV conditions are used for growth the process becomes very expensive and non scalable. Moreover transfer of the epitaxial graphene on arbitrary substrates is still challenging. Thus commercialization of this growth procedure is not feasible with current amount of knowledge and understanding.

## 2.2 Mechanical exfoliation

Although it has been known since the late 1970's that carbon precipitates in the form of thin graphitic layers on transition metal surfaces[1-2] no report on its electronic properties was made since the substrates were all metallic and conducting and no attempts on any technique to transfer them on insulating substrates were made. But in the late 90's Ruoff and co workers tried isolating thin graphitic flakes on SiO2 substrates by mechanical rubbing of patterned islands on HOPG[22]. However there was no report on their electrical property characterization. Using a similar method this was later achieved in 2005 by Kim and co workers and the electrical properties were reported[23]. None the less the first report of isolating graphene onto insulating SiO2 substrate by mechanical exfoliation was made by Geim and co-workers in 2004[4]. Since then various other techniques were developed to produce thin graphitic films and few layer graphene from bulk graphite or HOPG. Mechanical exfoliation is perhaps the most unusual and famous method for obtaining single layer graphene flakes on desired substrates. This method produces graphene flakes from HOPG by repeated peeling/exfoliation Fig. (1) a. This peeling/exfoliation can be done using a variety of agents like scotch tape[4], ultrasonication[51], electric field[52] and even by transfer printing technique[53-54] etc. In certain studies the HOPG has also been bonded to the substrate either by regular adhesives like epoxy resin[51, 55] or even by SAMs[56] to improve the yield of single and few layer graphene flakes. A recent study also demonstrates transfer printing of macroscopic graphene patterns from patterned HOPG using gold films[57]. It is by far the cheapest method to produce high quality graphene. Graphene flakes obtained by mechanical exfoliation methods are usually characterized by optical microscopy, Raman spectroscopy and AFM. One of the most popular methods of identifying single layer graphene is through optical microscopy. Depending on thickness graphene flakes give a characteristic color contrast on a thermally grown $SiO_2$ layer of 300nm thickness on top of Si wafers[58](Fig.2). Raman spectroscopy is also carried out on graphene obtained by mechanical exfoliation. It is the quickest and most precise method of identifying the thickness of graphene flakes and determining its crystalline quality. This is because graphene exhibits characteristic Raman spectra based on number of layers present[59-61]. AFM analysis is carried out on exfoliated graphene to verify its thickness and number of layers. However, finding a single layer flake is a matter of chance plus the yield of single and few layer graphene obtained by this method is very poor and the flakes are randomly distributed on the substrate. The graphene flakes obtained are very small and range from a few microns to a couple of millimeters in size. Thus this method is practicable only on a laboratory scale and is not scalable to a commercial level.

## 2.3 Wet Chemical synthesis

Chemical methods of producing colloidal suspensions of graphene and Chemically Modified Graphene (CMG) from graphite, derivatives of graphite and graphite intercalation compounds is another promising route for large scale graphene synthesis. Graphene and CMG produced by these chemical methods is suitable for a variety of applications such as paper like materials[62-67], polymer composites[13], energy storage materials[9] and transparent conductive electrodes[20]. Graphene has been mainly produced trough chemical route from graphite oxide. Graphite oxide was produced long ago in 1860 and since then it is being mainly produced by Brodie[68], Hummers[69] and Staudenmaier[70] methods. Each of these methods involves oxidation of graphite on presence of strong acids and oxidants. Graphite oxide is known to possess a layered structure of graphene oxide sheets and is highly hydrophilic. Thus it is easily intercalated with water[71]. However, the structure of graphite oxide has is still under debate in literature. Recent studies using solid state $^{13}$C labeled NMR on graphite oxide has thrown some light on its structure[72]. Another recent article on NMR study of graphite oxide has given further insight into its atomic structure[73]. Graphite oxide can be easily disintegrated into graphene oxide either by sonication[74] or stirring[75]. This colloidal suspension can be prepared either in water or in a variety of organic solvents[76]. This colloidal suspension of graphene oxide can be reduced using various methods to form a suspension graphene sheets Fig. (1) e. This, when dried is a black powder which is electrically conducting. However, this chemically reduced graphene oxide is not pristine graphene and still contains some amount of oxygen. Similarly, homogeneous colloidal suspensions of CMGs functionalized with numerous other groups in a variety of solvents have been reported[74, 76-79]. Some works also report CMG functionalization with nanoparticles like gold[80] and $TiO_2$[81].

Colloidal dispersions of graphene and CMGs can also be produced from other derivatives like graphite, graphite intercalation compounds[82] or expandable graphite. A homogeneous colloidal suspension of graphene sheets or ribbons can be obtained by stirring a potassium based graphite intercalation compound (K $(THF)_xC_{24}$) in N-methyl pyrolidone (NMP)[83]. Polymer coated graphene dispersions in organic solvents have also been successfully synthesized. Such dispersions have been mainly synthesized from expandable graphite[84-86]. An aqueous suspension of graphene derived from expandable graphite has also been reported. The graphene sheets thus produced are coated with organic molecules[87]. Electrochemical treatment of commercial graphite rods has also been done in a phase separated mixture of water and ionic liquid to produce ionic liquid functionalized graphene[88].

Colloidal suspensions of graphene and CMGs are very promising for preparation of many materials like transparent conducting films, paper like materials and graphene based composites. Recently colloidal graphene has also been directly exfoliated by carefully choosing the solvents and sonication[89-90]. Thereafter it has also been separated into different fraction according to layer thickness using Density Gradient Ultracentifugation[91-92]. It will also prove to be very promising for sensor applications[93]. However, due to the presence of large amounts of impurities like oxygen and also due to high defect density they are not suitable for many other electronic applications. Other looming issues are scalability and stability of the colloids. Finally, much better and thorough understanding of their structure and reaction mechanisms is needed for further developments in right directions. For a detailed discussion focused on chemically derived graphene and CMGs, see ref. [76] by Park and Ruoff.

**2.4 Epitaxial growth on carbides**

Epitaxial graphene growth on metal carbides has been practiced since long[94-101]. But none of the earlier reports have used Silicon carbide as substrate for epitaxial growth. Since the substrates being used were of conducting nature, no studies on the electrical properties of monolayer graphene were ever reported in the earlier works. However in 2004 a report on electrical measurements on patterned epitaxial graphene grown on SiC first appeared[102]. This is because SiC is a wide band gap semiconductor (3 eV) and thus electrical measurements can be carried out using it as the substrate. However, growing epitaxial graphene on SiC required UHV conditions and was thus difficult and expensive. Moreover, the quality of epitaxially grown graphene in SiC was not upto device levels and still posed many challenges. A significant breakthrough in overcoming this obstacle was achieved by growing high quality, large area epitaxial graphene at close o atmospheric pressures (900 mbar) in an argon atmosphere. The graphene films grown under argon also exhibit high crystalline quality, carrier mobility and Hall conductivity than graphene films produced by vacuum graphitization[103]. A few works have also reported the mechanisms and kinetics of epitaxial graphene growth on SiC[104-105]. A schematic of the synthesis process is shown in Fig. (1) c. It has been found that in vacuum graphitization the quality of films produced can be improved and the temperatures can be lowered by introducing a small vapor pressure of Si using disilane. This enabled in reducing the thermal etch pit formation and also enabled the growth of large area graphene flakes with uniform thickness. Another report gives concrete evidence using HRTEM that graphene nucleates along (1-10n) plane on the SiC surface steps and that the graphene quality improves with increasing temperature of graphitization[106]. Similar HRTEM studies have also been carried out on hexagonal face of 6-H SiC[107]. Epitaxial growth on SiC has also been achieved without subliming Si. This is done by an additional supply of carbon. This additional carbon may be supplied by hydrocarbon gas decomposition or by sublimation of solid carbon source in Molecular Beam Epitaxy[108]. Successful preparation of graphene monolayer on TaC, NbC, ZrC, HfC and TiC surfaces by ethylene decomposition has been achieved long back[95-96]. Epitaxial growth on SiC by MBE can be achieved at relatively lower temperatures than thermal graphitization. Another study reported epitaxial graphene on SiC at 950 C by evaporating cracked ethanol[109]. These techniques involving additional carbon evaporation prevents reconstruction of SiC surface which occurs in thermal graphitization due to higher temperatures and leads to uneven and rough surfaces. This phenomenon can also be prevented by employing these methods. The graphene grown epitaxially on carbides is again analysed using a variety of techniques like LEEM, LEED, RHEED, HREELS, ARPES as well as STM and AFM. The application of most of these techniques has been described above in Section 2.1 . Besides these HREELS is particularly used to study the phonon dispersion curves in epitaxial graphene[110]. This particularly helps in understanding the bonding/interaction of the grown graphene with the substrate. Epitaxial graphene growth on SiC has been visualized as a very promising method for large scale production and commercialization of graphene for applications into electronics. This is because since SiC is a wide band gap semiconductor, devices can be directly fabricated on it without the need of transfer. This makes the device fabrication very feasible and practical. However, there are still many obstacles to overcome for such a realization. A particular challenge from synthesis point of view is uniform growth of large area single layer graphene on wafer scale substrates. Another challenge is the improvement in quality of the graphene synthesized by this method.

**2.5 Chemical Vapor Deposition**

CVD growth of graphene on metallic thin films is very similar to epitaxial growth on metallic single crystals. However in CVD growth the metallic substrates are in the form of thin foils or films and the growth pressures are relatively higher than epitaxial growth. Moreover since the substrates are in the form of thin foils/films they are poly crystalline in nature and hence the individual grains/crystals are randomly oriented. Carbon segregation also occurs at grain boundaries which lead to folds or areas of higher thickness on the graphene films. CVD growth of graphene has been mainly practiced on copper[111-112] and nickel[113-115] substrates. It is interesting to know that the growth mechanism on each of these metals is different.

Nickel was the first substrate on which CVD growth of large area graphene was attempted. These efforts had begun right from 2007[116]. However better control of the growth procedure was not achieved and thus the works largely remained unnoticed. First concrete reports of controlled synthesis of large area graphene on nickel substrates appeared around late 2008[113] and early 2009[114]. These reports immediately spurred tremendous interests in CVD synthesis of graphene. Later many reports followed exploring growth mechanisms and optimizing growth conditions and parameters. A major drawback with the nickel substrate is that the growth process is not self limiting i.e the graphene film thickness is governed by the time of exposure to the carbon precursor as well as thickness of nickel film. Another disadvantage with nickel catalyzed growth is that a large number of wrinkles and folds are observed on the films. These folds or wrinkles form due to either due to difference in thermal expansion of graphene and nickel or due to defect nucleation at step edges on Ni surface[117]. The growth mechanism envisioned is similar to the epitaxial growth on Ni single crystals. In CVD growth each grain acts as a single crystal. Since the lattice mismatch of graphene is minimum with Ni (111) most favorable graphene growth occurs along the (111) surface. Thus the quality of graphene films can be significantly improved by producing a textured Ni film which can be achieved by specialized annealing treatments[118]. Studies on influence of cooling rates on graphene thickness and quality have also been done[119]. It has been found that very high cooling rates (100 C/s) do not produce any growth since all the carbon gets trapped within the nickel lattice and small cooling rates lead to formation of thick graphitic films. A moderate cooling rate of ~ 10-15 °C/s has been found appropriate for growth of mostly 1-2 layer films[114-115, 119].

However according to recent studies on CVD growth on copper have proved copper to be a more promising substrate[112]. This is because graphene growth on copper is self limiting i.e it stops after one layer growth. Moreover, high quality films comprising mostly of SLG can be produced on copper foils. The growth is also thickness independent and minimum surface finish is required for the substrate. The mechanism for graphene growth on copper is different than for nickel. The main reason being the wide difference in solid solubility of carbon between copper and nickel. While nickel has considerable solid solubility of carbon, copper has very limited solubility and thus carbon does not diffuse to a great depth in the copper substrate. A schematic diagram of growth process is shown in Fig. (1) d. Thus growth on copper is mainly a surface phenomenon and largely depends on carbon ad-atom concentration and vapor pressures of carbon source. However one disadvantage with copper substrate is that for synthesis of high quality single layer graphene films vacuum assisted CVD growth is required[112]. Thus low pressures of the order of 500 mtorr are essential to get high quality films. Graphene grown on copper foils as well as e-beam evaporated Cu thin films[120] shows no difference in crystalline quality and thus minimum surface preparation of the substrate is required. The crystalline quality of CVD synthesized graphene on copper foils using methane as a precursor was taken to a whole new level by optimizing the growth parameters. While it was shown that multilayer films grow using gaseous precursors at ambient pressures and high methane partial pressure, the use of ambient pressures and low concentrations of methane enabled monolayer synthesis thereby elucidating the different growth mechanisms on copper using Low Pressure (LP) and Ambient Pressure (AP) CVD and also the role of kinetic factors[121]. The Ruoff group successfully optimized all the growth conditions using a two step process to grow large domain size, continuous, single layer graphene films with very high mobility values in the range of 12000-16000 V cm$^{-1}$. The two step process involves use of high temperature and low methane partial pressure to enable low density nuclei formation, followed by high temperature high methane partial pressure for rapid and continuous single layer growth[122]. Using the same strategy and an even higher temperature (1035˚C) the same group demonstrated, large graphene single crystals upto 0.5 mm across grown on copper foils[123]. Another recent breakthrough in CVD synthesis of graphene was achieved by liquid precursor based high quality growth n large areas by Srivastava et al.[124]. In this work large area high quality graphene synthesis has been done using a liquid precursor hexane. Fig (3) shows a large centimeter size graphene film on Si/SiO$_2$ substrate. Using a liquid precursor introduces many advantages over conventional gaseous precursors and reduces severity of experimental conditions. Moreover, liquid precursors are easy to handle and may provide a breakthrough in substitutional chemical doping of graphene by boron or nitrogen since boron and nitrogen containing organic liquids are very easily and cheaply available. When it was already shown that CVD growth can be done using gaseous and liquid precursors, solid precursors shouldn't be left behind. This piece of the puzzle was completed by Tour and co workers by efficiently synthesizing large area graphene on copper foils using solid carbon precursors. The authors spin coated PMMA films (100nm) on copper foils and annealed these coated foils in Ar/H$_2$ atmosphere between 800 ˚C to 1000 ˚C. The authors also observed a control in the layer thickness by varying the flow rate of the Ar/H$_2$. Similar to liquid precursors, the solid precursors present a unique advantage of doping graphene with heteroatoms thereby allowing us to tune its electronic structure and properties. The authors also demonstrated this by spin coating melamine (C$_3$N$_6$H$_6$) +PMMA mixture on the copper foils and annealing. The resulting graphene shows N doped behavior and significant nitrogen was shown to be incorporated in the graphene lattice by XPS[125].

Since transfer of graphene films synthesized by CVD deteriorates the graphene quality as well as causes wrinkles formation, a transfer free direct device fabrication process has also been developed. This process involves selective etching of the underlying substrate by masking using photolithography and the remaining substrate can be used as electrodes. It was also observed that the e-beam evaporated Cu films must be greater than 500 nm thick for sufficient stability to facilitate uniform graphene growth[120]. CVD synthesis of graphene films on copper and nickel foils has been scaled up to wafer size production. They have also been successfully transferred to arbitrary substrates and their use as transparent conducting films has been demonstrated in flexible electronics and strain sensors[126]. Thus CVD synthesis of graphene on polycrystalline metal films is another very promising method for producing large scale graphene for electronics and other applications.

### 2.6 Miscellaneous methods

Besides the above mentioned methods a few alternative methods have also emerged for synthesis of graphene. Primary among them is the arc discharge method. Analogous to carbon nanotubes graphene can be synthesized by arc discharge of graphite electrodes. The synthesized graphene is in the form of nanosheets or nanoflakes which appears as black powder. Typically arc discharge is carried out under high pressure of hydrogen + helium for pure graphene[127]. There is even a report on graphene synthesis by arc discharge of graphite electrodes in air[128]. This process does not require any substrate or catalyst unlike other methods. It has been found that graphene formation occurs at very high gas pressures and at comparatively lower pressures, carbon nanotubes, nanohorns and other products is favored[128]. Arc discharge method is also very suitable for producing doped graphenes by introducing of nitrogen and boron containing gases in discharge atmospheres along with hydrogen[127] or even using ammonia or boron stuffed graphite electrodes[129].

Substrate free synthesis of graphene flakes and nanosheets using chemical vapor deposition is another alternative route. Many experimental reports of graphene synthesis have appeared in this area. Graphene flakes synthesized are in the form of a black powder[130-131]. A typical process of such kind involves aerosol formation of a carbon precursor followed by high temperature pyrolysis with the help of inert gas plasma to produce the graphene flakes. In some cases catalyst is also used to synthesize ribbon like graphene[132]. A solvothermal route has also been found for cheap and facile synthesis of graphene from non graphite precursor. This method promises large scale synthesis of graphene from cheap and easily available precursors like ethanol and sodium which are reacted to give an intermediate solid which is pyrolized and sonicated to give graphene dispersion[133].

However, for effectively putting graphene into applications high quality and large scale synthesis is essential. All the above mentioned techniques have their own advantages and limitations hindering their expansion to commercial levels. Thus clearly more focus is still needed on research to synthesize high quality and large area uniform graphene films by a cheap and facile route. For synthesis and detailed description of characterization techniques of graphene, readers are also advised to refer [134],[135-136].

## 3. Opening of band gap

Although graphene looked promising right from the first report for transistor applications and as a potential replacement for silicon, a major impediment for its application in the same has been its zero band gap. Due to this band structure graphene based FETs have low on off ratios. Thus one of the thrust areas in the field of graphene research right from the very beginning has been on techniques to open up a band gap in graphene without compromising on any of its other properties. Various techniques and methods have come up to do that. They will be covered in detail below.

### 3.1 Substrate induced band gap opening

This was the first technique to experimentally report band gap opening in graphene. STS/STM reports of band gap opening in epitaxial graphene on epitaxial h-BN have appeared in 2002 even before the first report of graphene isolation on insulating substrate. Epitaxial single layers of h-BN followed by graphene were synthesized on Ni (111). The graphene/h-BN/Ni(111) system showed a band gap of 0.5 eV in STS measurements thus indicating a substrate induced gap opening[137]. Recently theoretical studies have also been performed on h-BN substrate induced band gap opening in graphene. The DFT studies indicate that a band gap of 53 meV opens up in graphene on bulk h-BN. The study also reported band structure of epitaxial graphene on copper. It was found that a very small band gap close to zero opened up in graphene on copper[138]. Another recent theoretical study demonstrated that graphene deposited on oxygen terminated $SiO_2$ surfaces develop a band gap of about 0.52 eV whereas if the surface is OH terminated i.e hydrogenated, no covalent interaction occurs between graphene and oxygen and thus the semi metallic character of graphene is preserved[139]. More recently a report of band gap opening in epitaxial graphene on SiC due to interaction with the substrate has sparked a lot of interest because of the viability of the growth process. A schematic diagram for the same is shown in Fig. (4) a. However, epitaxial graphene on SiC is electron doped and the Fermi level lies above the gap. Thus to make graphene a viable semiconductor either it has to be hole doped or the Fermi level must be moved in between the gap by applying gate voltage[140].

### 3.2 Chemical substitution doping

Substitution doping is perhaps the most well known doping method as it is the primary method used in doping conventional semiconductors. In a graphitic carbon lattice, both n as well as p type of doping can be achieved by substituting carbon with nitrogen[111, 129, 141-148] and boron[129, 143-147, 149-152] or both[153-154]. Although a great deal of theoretical work has been done on substitutional doping of graphene, comparatively very few reports have appeared on experimental works verifying the theoretical results.

Nitrogen doping can convert graphene into a p-type semiconductor. A good amount of theoretical work has also been carried out to demonstrate this. It has been theoretically shown that N doping in ZGNRs energetically favors the ribbon edge. Moreover, two dopant atoms prefer to remain as far as possible from each other. Nitrogen doping has been experimentally achieved using CVD by mixing ammonia with methane as the precursor gas on a copper thin film substrate in a recent report which has generated considerable interest. A maximum of 8.9 % nitrogen doping was achieved. It also significantly improved the switching ratio in the fabricated FETs. However the carrier mobility was significantly reduced owing to the defects induced by doping[111]. A figure showing the doping sites and band structure can be seen in Fig. (4) b. Synthesis of doped graphene by arc discharge method has also been reported. Large quantities of N doped graphene have been synthesized by DC arc discharge using graphite electrodes in an ammonia atmosphere. Substitutional nitrogen doping occurs with nitrogen content in the sample not exceeding 1 wt% by XPS measurements[148]. Nitrogen doped graphene has also been similarly synthesized in a hydrogen+ (pyridine or ammonia) atmospheres with nitrogen content ~1% measured using XPS and EELS[129].

In the same way p doping of graphene can be achieved by substitutional doping of boron in the graphitic carbon lattice. A good number of theoretical studies on boron doped graphenes have already appeared. One of the earliest theoretical studies in this direction reports that presence of substitutional boron in the graphitic lattice improves the oxidation resistance of the graphitic carbon sample by changing the density distribution of high energy electrons. They also proposed that this oxidation inhibition was due to reduction in electron density and thus reduction in total number of active sites for carbon reaction. Their theoretical predictions were supported by experiments as well[149]. Recent studies however have focused more on electronic and transport properties of boron doped graphenes and nanoribbons. Boron doping leads to transformation of metallic zig zag nanoribbons to semiconductors. Moreover, similar to N doping it was observed that the substitutionally doped boron atoms also prefer the edge sites energetically. Also substitutionally boron doped GNRs can act as spin filters[151]. Another similar study shows that boron edge doped zig zag GNRs show half metallicity beyond a certain critical electric field thus having potential applications in future spintronic devices[150]. A theoretical study of substitutionally doped semiconducting armchair GNRs suggest that they can be metalized by alternating chains of 7 boron atom clusters (B7) embedded into their structure[152]. A recent theoretical study on large area doped graphene shows that the transport properties of large area graphene are not significantly affected even for doping levels as high as 4%. It also reports that minimum quantum interference effects are observed even for high doping values and that the carrier mobilities become more asymmetric with increasing doping. Holes are found to be more sensitive to localization phenomena than electrons in boron doping[143]. As far as experimental studies are concerned similar to nitrogen doped graphenes, boron doped graphenes have also been synthesized by arc discharge either in hydrogen + $B_2H_6$ (diborane) atmosphere or by using boron stuffed graphite electrodes. The boron content was found to be ~1-3% by using XPS and EELS[129]. Earlier, studies on boron doping in graphene layers of HOPG have also been done by high temperature diffusion in presence of $B_4C$. It was also confirmed by STM and Raman observations that substitutional doping of boron increases number of defects and disorder in the graphene layers[155]. However, very few reports of substitutional doping have appeared and uniform doping over large area graphene has not been successfully achieved as yet. Moreover, the carrier mobility and conductivity values of substitutionally doped graphene are far inferior to that of pristine graphene. None the less it is one of the most promising ways of opening a band gap in large area free standing graphene.

**3.3 Hybrids**

Theoretical studies have appeared since long on doping graphitic carbon lattice with both boron and nitrogen simultaneously forming BCN solid solutions in hexagonal lattice and synthesis of single layer BCN semiconductors and $B_x C_y N_z$ nanotubes [156-158]. There are some experimental works as well reporting successful synthesis of BCN materials [159-162]. BCN films have also been synthesized by CVD using C, B and N precursors. However, the thickness of these films has been of the order of a few nanometers to microns. Moreover, these films showed poor crystalline quality[161-163]. All these works report atomic hybrids of B, C and N with semiconducting electronic properties. However investigating into the B-C-N phase diagram it is found that boron and nitrogen have a tendency to segregate and form h-BN and is also energetically favorable[164]. Thus formation of graphene-h-BN hybrids is more feasible. Even this system has been theoretically investigated and has been recently synthesized experimentally. Hybrid C-BN nanotubes with BN domains have been synthesized earlier by laser vaporization however there was no report on their electronic properties[165]. Theoretical study on the same system has revealed band gap opening in such hybrid nanotubes. It was also reported that the N terminated BN domains were the most stable energetically[166]. Theoretical study of h-BN/GNR/h-BN sandwiched hybrid nanoribbon structures has also been recently reported. This system is structurally stable and that ZGNRs sandwiched between h-BNs behave as half metals and thus have potential applications in spintronics[167]. However, the recent synthesis of large area C-BN hybrids has generated tremendous interest in these materials. Large area 2-3 atomic layer hybrid films consisting of graphene and h-BN domains have been successfully synthesized on Cu substrates by CVD. These films were also synthesized over varying concentration ranges of C and BN. Small band gap opening has been observed and unique dual optical band gap was observed. These films had isotropic structure and thus the technique looks very promising for synthesizing hybrid films with tunable electronic and optical properties. Theoretical computations carried out in the same study also suggest band gap opening in hybrid films with BN domains and that the band gap increases with decreases in h-BN domain size[153]. This can be seen in Fig. (4) c. Although this method looks very promising, the control over domain size and shape has not been achieved yet which is essential for tuning the gap and other electronic properties.

**3.4 Quantum confinement (Graphene nanoribbons)**

As mentioned above, graphene has showed tremendous promise for applications in future nanoscale electronics and also as a potential replacement for silicon and even as conductive interconnects. However, its applicability cannot be effectively realized unless a finite band gap is opened up in its electronic structure. The above illustrated methods have been successful to quite an extent in opening up a finite band gap in graphene and thereby improving on-off ratios in graphene based switches. However a major drawback with most of the above methods is that the carrier mobility is tremendously reduced thereby affecting the device performance speed. Thus a method of opening finite band gap in graphene is desired which does not affect its carrier mobility to a great extent. This can be effectively realized by band gap opening through quantum confinement of electrons in graphene i.e graphene nanoribbon formation. In large area planar graphene the electrons are confined in 2 dimensions thus its band structure is different from three dimensional graphite. However by making a quasi 1 dimensional nanoribbon out of a large area 2-D graphene sheet would confine the electrons in a single dimension thereby expanding the electronic structure further and making GNRs finite band gap semiconductors similar to semiconducting SWNTs. The graphene ribbons were originally introduced as a theoretical model by Mitsutaka Fujita and co-authors to examine the edge and nanoscale size effect in graphene [168-170]. Since no impurities are being introduced in the graphitic lattice no charge scattering would occur except at the edges and thus carrier mobility will not be tremendously affected in a nanoribbon. Thus, graphene nanoribbons with width < 10 nm possess a finite band gap and also have enough carrier mobility thus facilitating high switching ratios[171](See Fig.(4) d). This was first demonstrated by Dai group at Stanford. Electronic confinement, coherence and Dirac nature of the carriers in graphene was demonstrated by de Heer and co workers in nanoribbon geometry [172] and also by Geim and co workers [173] in quantum dot geometry. So the all semiconducting property of sub 10 nm GNRs could then bypass the problem of extreme chirality dependence of electronic properties as observed in SWNTs and thus make GNRs an ideal candidate for nanoelectronics. Metallic GNRs are also ideal candidates as conductive interconnects than CNTs since they have a high current carrying capacity. However, controlled synthesis of GNRs in large quantities with narrow length and width distribution is very difficult at the moment. Various techniques for synthesizing GNRs are highlighted in the section 4 on nanoribbon synthesis.

**3.5 Miscellaneous methods of opening band gap:**

A few other methods also exist for band gap opening in graphene. A few works have focused on band gap opening due to nanohole superlattice in graphene. Theoretical calculations of the same system have also been carried out[174-178]. However a recent experimental demonstration of the same has generated significant interest in this area. This work reported creating a nanomesh from graphene using conventional block copolymer lithography. The necks formed between 2 holes act as nanoribbons with finite band gaps. Moreover by varying the hole diameter and density neck widths can be engineered to tune band gap. A high switching ratio ~100 was achieved at room temperatures. Fig. (5) shows a schematic and SEM image of a graphene nanomesh device. This approach is a promising development towards band gap engineering in graphene[179-180]. Another recent report which has generated considerable interest is band gap opening in graphene by selective hydrogenation. Epitaxial graphene on iridium was subjected to hydrogenation by exposure to atomic hydrogen and the carbon atoms on the moiré fringes were selectively hydrogenated. ARPES was used to measure a band gap opening of 0.73 eV for 23% area covered by hydrogenated carbon. STM analysis was also carried out to verify the result. DFT calculations using tight binding approximation were also carried out to confirm the experimental observations[181]. A recent report of amorphous 2D carbon material which is random network of polygons (namely with 5,6,7 and 8 sides) in combination with vacancies, divacancies, dislocations and Stone Wales defects, which was essentially created by bombarding graphene with high energy electron beam in a TEM is another interesting prospect. The authors observed that graphene turns into a 2D amorphous material rather than a porous and defective crystalline structure and the DFT calculations indicate that this random network of carbon polygons can actually possess a finite band gap of the order of 200 meV which is also an underestimate under the GGA approximation used for the DOS calculations[182]. However, the large scale applicability and experimental proof of the gap opening remains an open question.

Molecular doping and charge transfer methods have also generated considerable interest in recent times to modulate the electronic properties of graphene. A change in the band structure occurs and carrier mobility becomes asymmetric as well. Molecular doping has been extensively studied both theoretically as well as experimentally. Geim and co workers have shown the effect of molecular doping by paramagnetic and diamagnetic molecules on the Hall conductivity of graphene. It was proved that diamagnetic adsorbates/impurities act as weak dopants because of the completely filled/closed electronic shells thereby exhibiting inert behavior and thus the inability to dope graphene. Whereas, paramagnetic adsorbates and impurities can effectively dope graphene due to their open electron shell configuration. Thus graphene can be effectively doped by any adsorbate even if there is a small mismatch in the chemical potentials of graphene and the dopant molecule[183].

Selective chemical fuctionalization of large area graphene in order to perturb its electronic structure for band gap opening in graphene also remains an interesting prospect. The first experimental demonstration of this kind of gap engineering was provided by Geim and co workers[184] by completely hydrogenating graphene in an atomic hydrogen atmosphere to form a complete insulator, graphane. They also observed that this hydrogenation reaction can be completely irreversible thereby enabling an extra degree of freedom in terms of gap engineering. Based on same concept many other theoretical studies have been performed to demonstrate tunable band gap and spintronic properties in graphene-graphane superlattices[185-187]. On similar lines other notable experimental demonstrations of opening band gap by chemical functionalization in graphene have been using Fluorine. Fluorographene was produced by exposing pristine graphene to Xenon Difluoride ($XeF_2$). The authors observed that fluorination doesn't significantly affect the mechanical properties of graphene as it still retains a Young's modulus of 100 N m$^{-1}$ and undergoes strain of 15% before failure. It is highly insulating and has an optical gap of 3 eV as well as stable in air upto 400 °C[188]. Another interesting study on graphene fluorination reports controlled fluorination of graphene on single and both sides of the membrane. They observed that while single side fluorination saturates at 25% coverage indicating a $C_4F$ structure by DFT calculations and band gap of 2.93 eV, double sided fluorination leads to a 50% coverage indicating a CF structure with a calculated band gap of 3.03 eV. They also observe that fluorinated graphene renders it optically transparent thereby indicating its insulating nature [189].

## 4. Nanoribbon synthesis

Considering the fact that quantum confinement can open up a significant band gap in graphene so as to facilitate high switching ratios in GNR based FETs a gold rush began to experimentally synthesize graphene nanoribbons and demonstrate the property. Two of the first successful reports in this direction was by Kim and coworkers in mid 2007[190] and by Dai and coworkers in early 2008[171]. Kim and co workers fabricated nanoribbons using e-beam lithography however they were not able to reach a width below 20 nm. Nonetheless the study was pioneering in demonstrating the effect of quantum confinement on electronic properties of graphene. In the other study graphene nanoribbons were collected in extremely small quantities by thermal exfoliation and sonication of HOPG in polymer solvent followed by centrifugation to separate large HOPG pieces and leave behind the nanoribbons. The electrical measurements performed on these nanoribbons showed that a significant band gap had opened in nanoribbons with width <10nm and showed high switching ratios > 10$^5$ even under a bias of 1 V. Following these report the nanoribbon gold rush began. Described below are various techniques which have been used to produce graphene nanoribbons.

### 4.1 Patterning and etching

One of the first and foremost techniques which enabled synthesis of graphene nanoribbons was e-beam lithography. The pioneering work in this direction was carried out by Kim and co workers. Mechanically exfoliated single layer graphene flake was patterned in the form of nanoribbons of various widths using Hydrogen Silsesquioxane (HSQ) which is negative electron resist. After developing the exposed areas were etched off by exposing the patterned film to oxygen plasma. Electronic measurements were carried out on the nanoribbons and a finite band gap as observed in these ribbons which was inversely proportional to the ribbon width in agreement with theoretical predictions. However, on effect of band gap on crystallographic orientations of the ribbons was observed in contrast to theoretical predictions of extreme chirality dependence of electronic properties. This difference was attributed to rough edge structures which were generated by the oxygen plasma etching process and thus precise control over the atomic structure at edges was not possible by etching technique[190]. A similar work was carried out around the same time by Avouris and co workers[191]. Besides, e-beam lithography there have been other attempts of generating graphene nanoribbons from large area graphene. Notable among them is etching using nanowire mask. In this method silicon nanowires grown using Au catalyst by V-L-S technique are dispersed over macroscopic graphene obtained my mechanical exfoliation or chemical techniques in a silicon substrate. Following this the graphene is exposed to oxygen plasma. Thus, the exposed area is etched away by oxygen plasma while the regions coved by nanowires are not etched and thus form nanoribbon structures. The width of the nanoribbons formed can be controlled by the diameter of nanowires and also by time of exposure to oxygen plasma. Higher the time of exposure smaller is the width of the nanoribbons. After etching the nanowires can be easily removed by brief sonication of the substrate leaving behind the nanoribbons on silicon substrate[192]. The same route has been used to fabricate GNR transistors with high K gate dielectrics which have reported the highest transconductance values of GNR transistors till date. In this process hafnium oxide (high K dielectric) coated Si nanowires were used as a mask for etching graphene. A thin amorphous layer of hafnium oxide is deposited onto Si nanowires by ALD[193]. Another recent work reports precise fabrication of graphene in sub-20 nm regime using a helium ion beam. This work was carried out with the help of an helium ion microscope and look promising for graphene fabrication and lithography in future[194]. Patterning and etching of graphene to produce nanoribbons saw a major breakthrough achieved by Tour and coworkers wherein they demonstrated nanoscale patterning of various forms of graphene by zinc sputtering and dissolution. The authors report that sputter depositing zinc on graphene removes one layer of graphene at a time thereby enabling them to precisely control the device geometry one atomic layer at a time. This discovery come from the fact that, sputter depositing zinc damages the top layer of carbon atoms also partially embedding it in zinc lattice and goes that layer away along with zinc in the dissolution process. They demonstrated this with four different kinds of graphene including CVD grown graphene and chemically derived graphene thus holding promise for large scale applications. Further this single layer etching only happens with moderately reactive metals like zinc and aluminum and not with gold and copper which can be used as interconnects and thus opens up new avenues for fabricating novel graphene devices with controlled number of layers and geometry[195]. However, the current ways of fabricating GNRs by patterning and etching do not allow precise control of nanoribbon width. More importantly the nanoribbon edges are extremely rough leading to the loss of chirality dependent electronic and spintronic properties. Thus, GNRs fabricated by this route are not suitable for spintronic applications. Fabrication of very thin nanoribbons (<8 nm) is also very difficult and has not been effectively realized.

### 4.2 Catalytic nanocutting of graphene

In order to achieve precise control over edge structure of graphene nanoribbons, catalytic nanocutting of graphene is a very promising method. In this method metal nanoparticles are used as catalysts which etch out the graphene along specific crystallographic orientations via catalytic hydrogenation process to produce atomically smooth graphene islands and nanoribbons. The edge structure of the cutting depends on the size of catalyst nanoparticle and the cutting occurs along straight lines unless the particle encounters a defect or an edge. Pioneering work in nanocutting was carried out by Ci et al.[196] and Datta et al.[197] These works have reported cutting using nickel and iron nanoparticles respectively. Following these works many other works reporting cutting using various other catalysts have been reported[198-201]. The atomic precision of edges, smoothness of cut and orientation of the cutting has been determined by TEM[199-201] and STM[196] studies. Etching of graphitic layers along specific crystallographic directions by catalytic hydrogenation has been studied earlier[202-205]. However, its applicability in controlled shape engineering of graphene appears very promising. Nanocutting can be achieved either on HOPG or even on graphene deposited on silicon substrate directly so that devices can be easily fabricated and integrated. Controlled cutting may also open up a possibility of producing 100% graphene circuits in which semiconducting and metallic nanoribbons are directly fabricated in the form of an interconnected network. Its mechanism is exactly the reverse of carbon nanotube growth. While carbon dissolves in catalyst nanoparticles to come out as CNTs, it gets dissolved and reacts with hydrogen to produce methane and move ahead to form the cut. The only difference is carbon concentration in the atmospheres. CNT growth requires a high concentration of carbon in the growth atmosphere while nanocutting requires very low concentration of carbon. As mentioned above the orientation of cutting depends on particle size. It has been observed that cutting mainly occurs along [1$\bar{1}$00] for particles with diameter > 10 nm while it proceeds mostly along [11$\bar{2}$0] direction for particles less than 10 nm. The cutting direction is preserved in most cases whenever the particle deflects and changes the cutting direction due to a defect or edge in case of graphene. This is in contrast to the case of graphite where cutting orientation is not preserved in 35% of cases. Thus 60° and 120° are the most common angles observed in crystallographic etching of graphene while 30°, 90° and 120° are rarely observed[206]. However, the major limitation with catalytic nanocutting is that even though control on edge structure and smoothness has been considerably achieved, there is still little control over shape and size of the islands or nanoribbons produced. Even though it is feasible to cut out GNRs with atomically smooth edges we still cannot control the GNR width and location. This puts a major limitation on its wide scale applicability. Moreover, since multiple cuttings are produced with a single particle they are very closely spaced so device fabrication is very difficult. This is one of the main reasons that no reports of electrical measurements on GNRs fabricated by catalytic nanocutting have appeared. Control of shape and size of features can be achieved to some extent by precise engineering of defects and step edges where nanoparticles can turn or stop. A report on such attempt to control cutting dimensions has already appeared[51]. The HOPG surface is selectively patterned and etched by conventional optical lithography and then the catalyst is deposited only in the trenches so that cutting occurs only on the raised platforms and its size is limited to the platform width. The same work also reports direct transfer of the nanocutting from HOPG to silicon substrate. But still in spite of all these attempts a much better control over cutting direction, dimensions and location is desired in catalytic nanocutting processes in order to realize its promise for GNR fabrication. A combination of nanolithography and catalytic nanocutting may provide a viable solution to address the above problems.

**4.3 SPM based lithography**

Nanometer scale lithography using STM has been demonstrated way back in 1985[207]. HOPG is a favorable material for STM experiments in air and early attempts to modify HOPG surface using STM tip were witnessed. In these early studies it was reported that nanometer scale lithograph on HOPG is possible only under presence of water vapor however later it was established that decomposition of water vapor takes place locally in the region of tunneling current leading to oxidation of the HOPG carbon thus forming nanometer size features and etchpits[208]. Thus with careful tuning of the tip velocity and applied bias GNR fabrication with one layer thickness is possible and has been recently demonstrated[209]. Using this technique sub 10 nm width GNRs were fabricated from HOPG using STM lithography. The HOPG surface was first imaged and then rotated so as to orient along particular direction and then was etched by applying higher bias than what is used for imaging. Then using STS the electronic structure of the GNRs was determined. A gap of 0.18 eV was observed for 10 nm wide armchair GNR which is in good agreement with theory. GNRs upto 2.5 nm were fabricated and a gap of 0.5 eV was measured for the same. Another work by the same group shows precise manipulation of singe atomic layers in graphene to produce trenches, cuts, holes, folds and ribbons of controlled dimensions and depths[210].

Nanometer scale manipulation of graphene and HOPG surfaces has also been recently reported using conductive AFM. The cutting mechanism is again electrochemical in nature similar to the case of STL as mentioned above. A conducting doped silicon tip is used for the lithography and the sample is placed on a heavily doped $Si/SiO_2$ substrate. A bias of about 15-30 V is normally applied so as to carry out the electrochemical oxidation. This mechanism has also been widely referred to as Local Anodic Oxidation (LAO). Humidity control is important and a humidity of around 55-60 % is essential for the stability of water meniscus. This water meniscus acts as an electrolyte for the electrochemical oxidation of carbon. Nanoribbons of approximately 25 nm width have been successfully fabricated using this technique [211-212].

The main advantage with the SPM lithographic techniques is the small resolution they provide and the specific atomically sharp edge control which can be particularly achieved in case of STL. Another major advantage with SPM nanofabrication processes is that they do not involve the use of any resist and thus involve minimum contamination. Moreover, the above mentioned processes are also carried out under ambient atmospheres and thus do not need controlled environments. However, scaling these SPM processes for large volume fabrication of nanodevices form graphene is rather unviable at the moment due to its slow speed of operation. None the less with concepts like Millipede[213] and Dip Pen Nanolithography (DPN)[214] already developed which uses thousands of tips simultaneously to allow massive parallel STL processes, such impediments to large scale nanofabrication using SPM processes may be eliminated. For a detailed article with special focus on SPM based lithography techniques of graphene see ref [215].

**4.4 Unzipping of carbon nanotubes**

Since carbon nanotubes can be thought of as rolled up seamless cylinders of graphene, unrolling or opening up nanotubes is a good way to produce planar graphene. This breakthrough was recently achieved by Dai et al.[216] and Tour et al.[217] These groups were successful in opening up multiwalled carbon nanotubes to form graphene nanoribbons. The width of the nanoribbons thus produced depends on the diameter of the precursor nanotubes. This unzipping of nanotubes was achieved by two entirely different methods. Dai et al. first embedded multiwalled CNTs in a PMMA film and then exposed this film to Ar plasma for various times so as to etch out top exposed portions of tube walls thereby opening up the tubes. With increasing amount of exposure times GNRs with reduced thickness i.e lesser number of layers were observed. In contrast to this approach, Tour et al. opened up the MWCNTs by oxidizing them in presence of sulphuric acid and potassium permanganate. It was observed that the severity of oxidation increased with the concentration of $KMnO_4$ and this led to oxidation of more layers of MWCNTs. The resulting nanoribbons which were produced were of graphene oxide and contained a lot of hydroxyl and epoxy groups which was verifies using ATR IR spectroscopy, XPS and TGA. However, they could be reduced back to graphene using hydrazine and high temperature hydrogen annealing.

Following these pioneering works various other attempts were made to synthesize GNRs by unwrapping CNTs. Catalytic nanocutting of multiwalled carbon nanotubes is one such method. Similar to the case of nanocutting of large area graphene using metal nanoparticle catalysts as described above in the section of catalytic nanocutting (4.2) above, nanotubes can also be longitudinally unzipped using this very technique. In this method nanotubes were first dispersed in a metal salt solution and the mixture was dropped and allowed to dry on a substrate. Following this the substrate was annealed in $Ar/H_2$ mixture at 500°C to allow nucleation of the nanoparticles followed by nanocutting at elevated temperatures of 850˚C[218]. Cuts of varying lengths and depths were produced. This technique allowed production of partial or completely unzipped carbon nanotubes. The main advantage with this technique as mentioned above is that it will produce atomically smooth and sharp edges GNRs with specific orientations. It can also be scaled for large volume production. However, the major problem with this technique is that it does not ensure complete cutting in all MWCNTs. There may be many nanotubes will be just partially cut and thus it puts a major limitation on its applicability.

Another interesting method for unzipping MWCNTs to GNRs is using electric field. In this in situ TEM study an electric field was applied to a single MWNT using a tungsten electrode and it was observed that the non contact end of the MWCNT started unwrapping and forming graphene nanoribbon. This GNR thus formed slides over the remaining MWCNT core and comes out. The electrical breakdown behavior of this GNR was also studied. It was found that a GNR undergoes a catastrophic electrical breakdown under high bias as opposed to step wise breakdown of MWCNTs. Another interesting observation was that the electrical conductivity of the GNR can be maintained even under severe mechanical deformations[219]. Production of graphene flakes from MWCNTs by using high DC pulses has also been recently reported[220]. This suggests that electrical unwrapping of MWCNTs could prove to be promising for large scale GNR production. This is the only unzipping technique which produces nanoribbons without any chemical or catalyst contamination and thus is viable for high purity GNR production for electronic applications.

Other miscellaneous methods have also been successfully employed for unzipping nanotubes but due to the major limitations may not be promising for successful production of GNRs on a large scale. Terrones and co workers developed a technique to open up MWCNTs by intercalation of lithium[221-222]. Another technique involves opening up of MWCNTs longitudinally by tunneling current of STM[223]. It opens up by a similar mechanism as proposed in STM lithography method mentioned above however in this case the MWCNTs have been functionalized prior to unzipping using STM tunneling current. The MWCNTs were functionalized using a cycloaddition reaction and then unzipped by in situ STM manipulation. The unzipping was achieved by interaction of the cycloadduct with energy supplied by STM and thus did not involve oxidation of graphene edges. Moreover, it is a clean reaction environment and does not involve any contaminants[223]. Another recent work by Dai and co workers demonstrates a facile method for synthesis of large quantity of GNRs. The MWCNTs are partially oxidized and then sonicated so that they open up into graphene nanoribbons. The nanoribbon quality and yield is higher and better compared to the plasma etching method by the same group reported earlier[224]. For detailed perspective focused on unzipping of CNTs see ref [222] by Mauricio Terrones.

Unzipping of CNTs to produce GNRs is certainly promising and may be scaled up for large scale GNR synthesis however, it faces the same old challenge which is associated with CNTs, i.e polydispersity. If the precursor CNTs are polydisperse the GNRs thus produced by unzipping them will also be of varying widths, thickness and chiralities. So again producing GNRs with controlled width and chiralities is an issue with CNT unzipping methods. Moreover, in most of the unzipping methods the GNRs produced do not have atomically smooth edges so their applicability for electronic devices gets limited. In many of the above mentioned unzipping methods a lot of defects are also introduced during unzipping thereby severely affecting electron mobilities further limiting the applicability of these GNRs in electronics. There is also a lot of contamination of the resulting GNRs induced by many of these techniques. So still there are a lot of challenges and hurdles to be tackled before CNT unzipping becomes viable for mass production of GNRs. Nonetheless most of the above mentioned problems can be addressed if unzipping of monodisperse SWCNTs. However no such studies have been reported as yet and producing monodisperse SWCNTs is a challenge in itself though numerous developments have been made in the same.

**4.5 Miscellaneous methods:**

Since graphene nanoribbons can be essentially considered as macromolecules of carbon or pure carbon polymeric chains, it should be possible in principle to synthesize them using preferential and selective self assembly of planar hydrocarbons in one dimension. This is a bottom up approach to GNR synthesis and was first demonstrated very recently by Fasel and co workers in a breakthrough paper. The authors used a brominated hydrocarbon which self assembles on a gold surface. The first step in the reaction is dehalogenation which occurs at 200˚C and which leads to a hydrogenated polymer chain on the surface. The next step involves dehydrogenation at 400˚C to produce pristine GNRs with atomically precise width and edge defined by the molecular precursors. The authors also demonstrated that using different shapes of molecular precursors halogenated at specific sites using nanoribbon and one dimensional graphene nanostructures can be synthesized from a bottom up approach on a large scale[225]. However since this reaction is carried out on gold no, electrical measurements on these unique 1D GNR structures was possible. Therefore an alternative synthesis route must be developed on an insulating substrate to realize its potential applications and properties.

Another important development in GNR synthesis was demonstrated by de Heer and co workers by growing GNRs in a templated fashion on SiC substrate. The authors exploit the selective nature of certain crystallographic facet of SiC substrate, namely (1-10n) to graphitize first due to weaker bonding of silicon atoms. Such steps were deliberately fabricated over the entire surface by lithography and fluorine based reactive ion etching to produce nanofacets of desired orientation and height. Later the sample was first annealed at 1200 ˚C -1300 ˚C for 30 min followed by graphitization at 1400 ˚C for 1.5 min. The result was formation of GNR with width equal to step heights all over the surface. GNR as narrow as 40 nm were successfully fabricated with mobilities of 2700 V cm$^{-1}$ at room temperature and on off ratio of 10 at low temperature (4 K) indicating band gap opening due to quantum confinement. This technique even allows large scale fabrication which the authors demonstrated by fabricating 10000 top gated GNR transistors on 0.24 cm$^2$ SiC chip[226].

## 5. Conclusions and future outlook

Considering the progress which has happed over the last 3-4 years in graphene research, this single atom thick material does look promising for a wide range of applications. However, there are many challenges yet to overcome. Synthesizing large area high quality single layer graphene is one of the major challenges which was pointed out in the beginning of this article. Recent developments, particularly in CVD synthesis of graphene on metal substrates have proved to be major breakthroughs in overcoming this challenge. This means large graphene based devices like flexible touch screens and transparent conductive panels should get commercialized in near future as roll to roll production of high quality graphene has already been demonstrated[227]. But still the synthesis of large area single domain graphene sheets at low temperatures remains a challenge. Currently the graphene synthesized by CVD route has multiple domains and thus the domain boundaries hamper the carrier mobility thereby making it unsuitable for device applications which require defect free, single domain graphene. The polycrystalline nature of the growth substrates also affects the graphene quality in that the graphene thickness is different on grains of different orientation and is abruptly thick with wrinkles at the grain boundaries. Thus obtaining high quality graphene for fundamental studies is still limited to mechanical exfoliation method. Another impediment for realizing graphene based electronics is opening up of band gap without affecting other electronic properties. The most viable solution in this direction is by quantum confinement i.e nanoribbon formation. However the persisting issues of polydispersity and dimension control come into picture with nanoribbon formation. Thus a major breakthrough is desired in synthesis of GNRs with controlled width and chirality. Since GNRs are direct band gap semiconductors, large quantities of monodisperse GNRs can be used for a variety of semiconductor based applications. Once these problems are addressed graphene based electronics should become a reality in the future. Substitutional doping of graphene is another promising method which can open up the graphene band structure without significantly affecting other properties. Although some amount of work has been done and results have proved to be promising, still uniform doping over large area graphene has not been effectively realized. Another way around the band gap opening problem in large area graphene would be to have an ordered functionalized superstructure formed by selective functionalization of large area graphene. This will create a periodic potential over the entire graphitic lattice thereby breaking the symmetry between the two sub-lattices and opening a band gap. This may be possible to achieve by forming a nanoporous ordered self assembled monolayer on graphene which followed by the desired chemical reaction which will take place only at the exposed carbon atoms while not at the others masked by the self assembled monolayer. A breakthrough method of producing large area doped graphene with non zero and tunable band gap can pave the way for graphene based solar cells which can potentially have very high efficiency compared to current silicon based photovoltaics. Another important area where graphene based applications are yet to be realized experimentally is spintronics. A number of theoretical studies have predicted a variety of spintronic properties and applications of graphene however none have been realized yet experimentally. For spintronic applications graphene with a uniform and atomically smooth edge is desired which is difficult to synthesize. Although a few successful attempts have been made in this direction by catalytic nanocutting method, still specific edge and width control at the desired location has not been achieved. Thus a large number of challenges are yet to be overcome to enable graphene based electronic applications; however rapid progress is being made in every direction which promises to overcome these problems in the near future.

## Acknowledgments

The authors are extremely thankful to Mr. Kanhaiya Prajapati and Mr. Gaurav Jain for assistance with the figures and graphics. AS acknowledges the support from Department of Science and Technology (DST), GOI, under BOYSCAST fellowship and CAS program at Department of Physics, B.H.U., sponsored by UGC, India.

Figures:

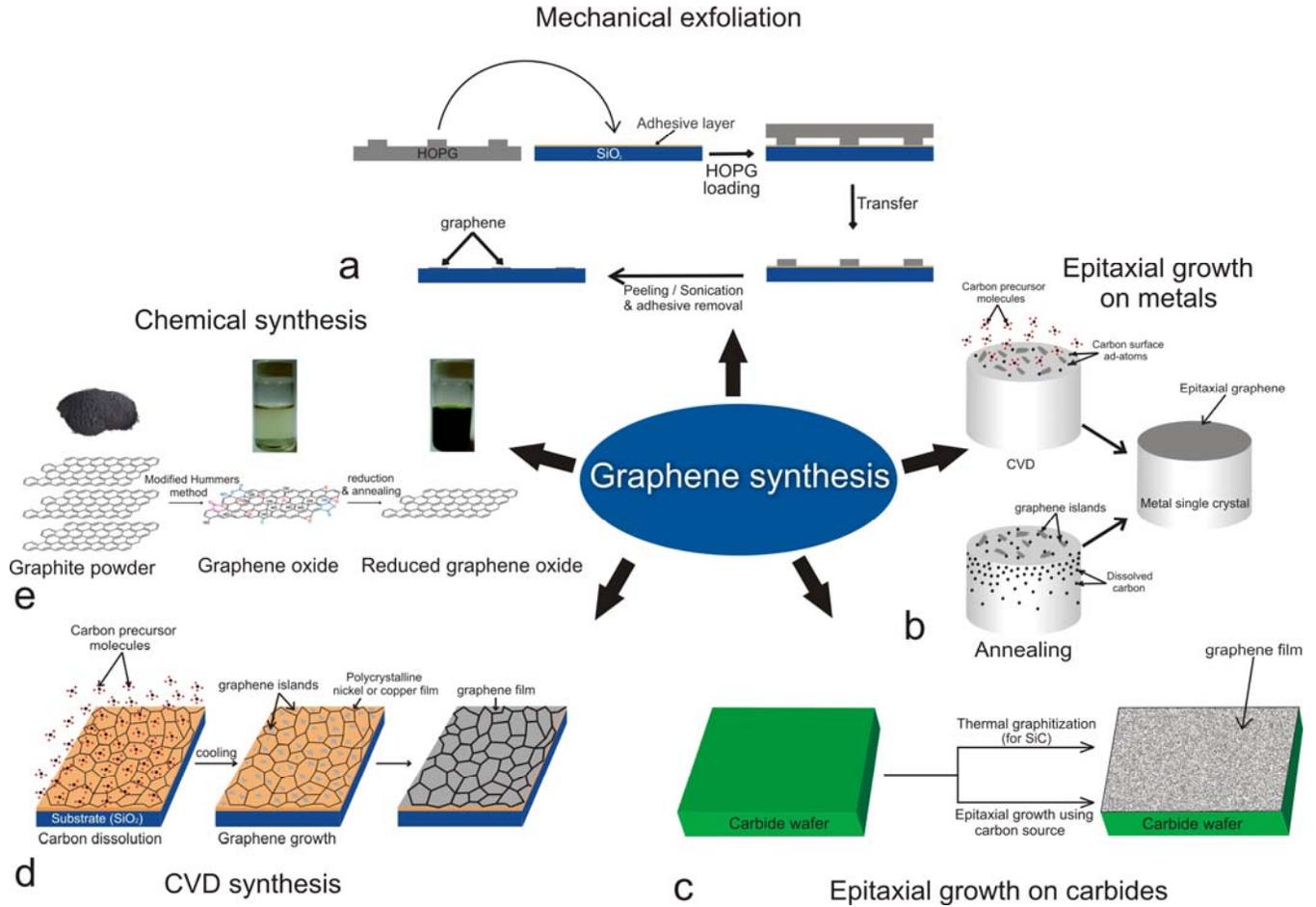

Figure.1: Schematic of various graphene synthesis techniques. (a) Graphene flake synthesis by mechanical exfoliation form HOPG. (b) Epitaxial graphene growth on metal single crystals. (c) Epitaxial growth of graphene on carbide wafers by graphitization or deposition. (d) CVD synthesis of large area graphene on metal (Ni or Cu) foils. (e) Chemical synthesis of graphene from graphite oxide. [Reprinted with permission from the Macmillan Publishers Ltd.[Nature Chemistry]; Ref. 73: W. Gao, L. B. Alemany, L. Ci and P. M. Ajayan, Nat Chem 1 (5), 403 (2009), Copyright (2009)]

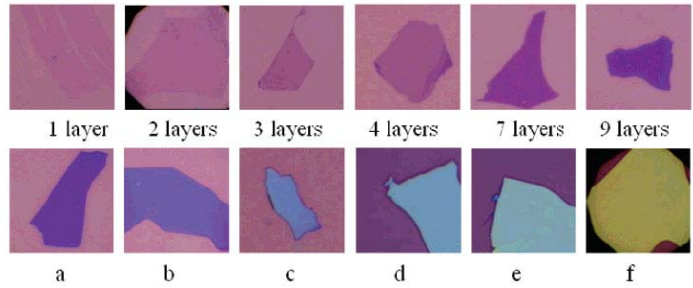

Figure.2: Graphene flakes of varying thickness on 300nm SiO$_2$/Si substrates [ Reprinted with permission from the; Ref. 58: Z. H. Ni, H. M. Wang, J. Kasim, H. M. Fan, T. Yu, Y. H. Wu, Y. P. Feng and Z. X. Shen, Nano Letters 7 (9), 2758 (2007); Copyright (2007), American Chemical Society ]

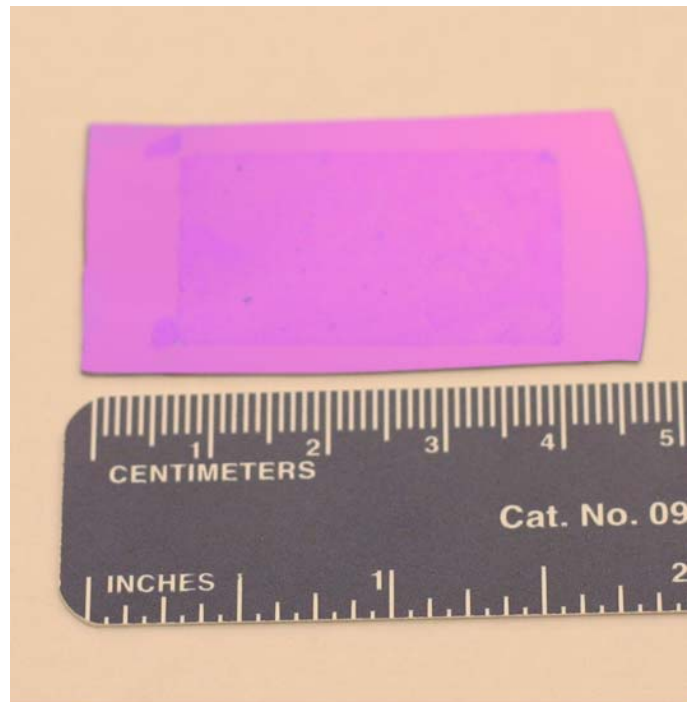

Figure.3: Large area, continuous graphene film on Si/SiO$_2$ substrate grown by CVD route on copper foils using liquid precursor. [ Reprinted with permission from Ref. 124: A. Srivastava, C. Galande, L. Ci, L. Song, C. Rai, D. Jariwala, K. F. Kelly and P. M. Ajayan, Chem Mater, (2010); Copyright (2010), American Chemical Society]

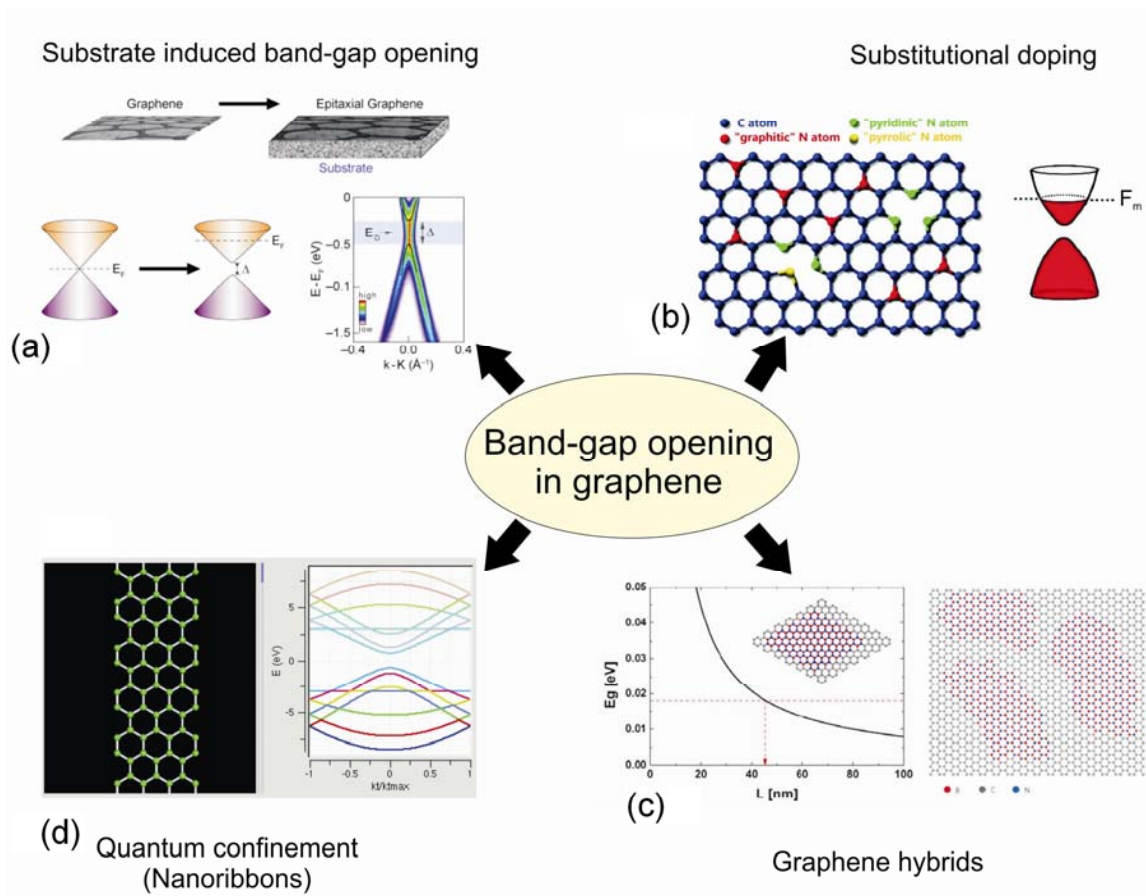

Figure 4: Schematic of various methods of opening band gap in graphene. (a) Substrate induced band gap opening in graphene on SiC substrate. [ Reprinted with permission from the Macmillan Publishers Ltd.[Nature Chemistry]; Ref. 140: S. Y. Zhou, G. H. Gweon, A. V. Fedorov, P. N. First, W. A. de Heer, D. H. Lee, F. Guinea, A. H. Castro Neto and A. Lanzara, Nat Mater 6 (10), 770 (2007); Copyright (2007)] (b) Band gap opening by substitutional doping of nitrogen in graphene lattice. . [ Reprinted with permission from Ref. 111: D. Wei, Y. Liu, Y. Wang, H. Zhang, L. Huang and G. Yu, Nano Letters 9 (5), 1752 (2009); Copyright (2009), American Chemical Society] (c) Band gap opening in graphene h-BN hybrids. . [ Reprinted with permission from the Macmillan Publishers Ltd.[Nature Materials]; Ref.153: L. Ci, L. Song, C. Jin, D. Jariwala, D. Wu, Y. Li, A. Srivastava, Z. F. Wang, K. Storr, L. Balicas, F. Liu and P. M. Ajayan, Nat Mater 9 (5), 430 (2010); Copyright (2010)] (d) Band gap opening due to quantum confinement in graphene nanoribbons.

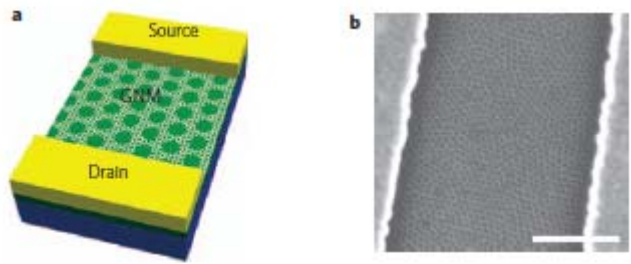

Figure.5: (a) Schematic diagram of a graphene nanomesh device. (b) SEM image of the device. [ Reprinted with permission from the Macmillan Publishers Ltd.[Nature Nanotechnology]; Ref. 178: J. Bai, X. Zhong, S. Jiang, Y. Huang and X. Duan, Nat Nano 5 (3), 190 (2010); Copyright (2010)]

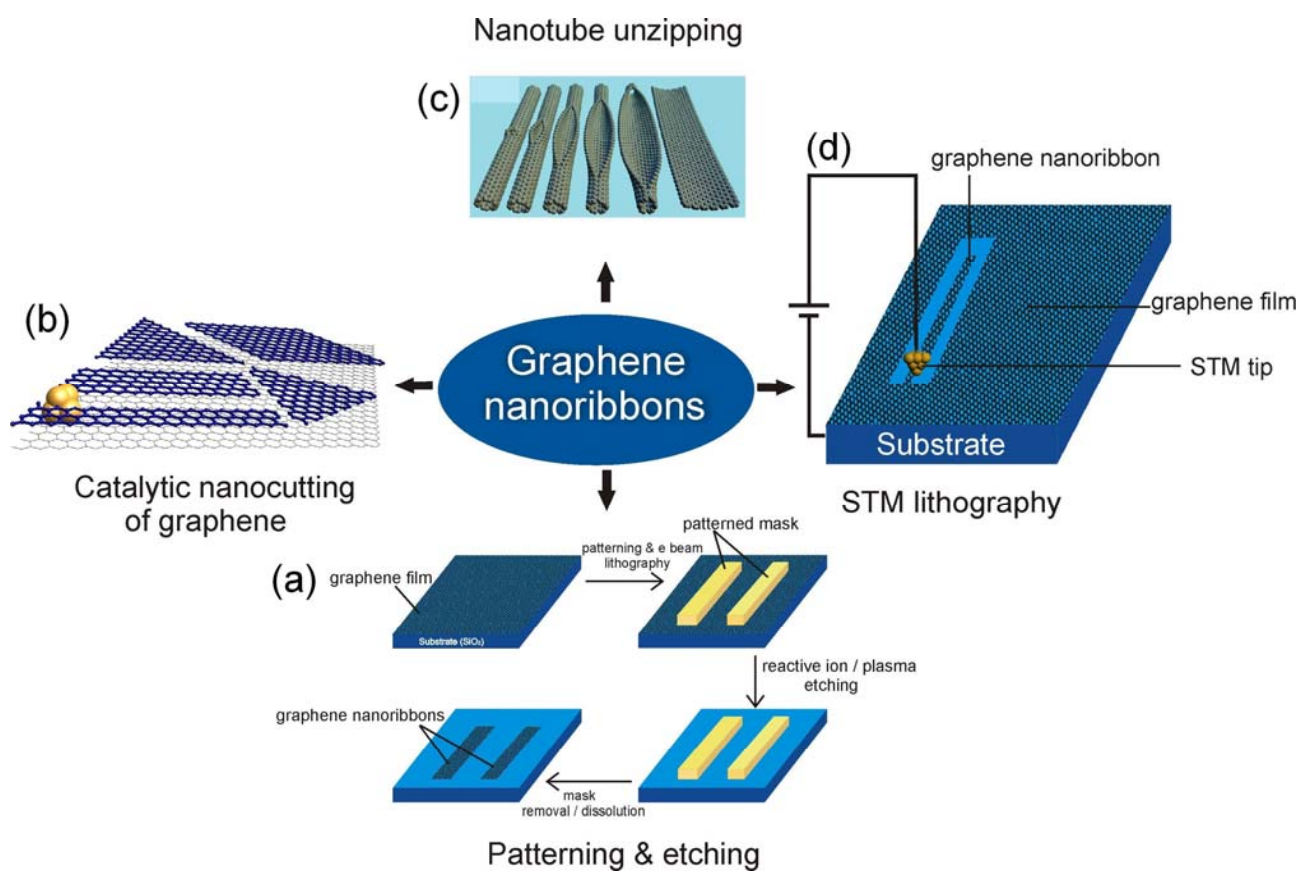

Figure 6: Schematic of various nanoribbon synthesis techniques. (a) Nanoribbon fabrication by patterning using e-beam lithography and oxygen plasma etching. (b) Catalytic nanocutting of graphene using metal nanoparticle catalysts to etch out graphene nanoribbons. [Reprinted with permission from the Wiley VCH; Ref. 51: L. J. Ci, L. Song, D. Jariwala, A. L. Elias, W. Gao, M. Terrones and P. M. Ajayan, Adv Mater 21 (44), 4487 (2009)] (c) Longitudinal unzipping of carbon nanotubes to form graphene nanoribbons. [Reprinted with permission from the Macmillan Publishers Ltd.[Nature]; Ref. 216: D. V. Kosynkin, A. L. Higginbotham, A. Sinitskii, J. R. Lomeda, A. Dimiev, B. K. Price and J. M. Tour, Nature 458 (7240), 872 (2009), Copyright (2009)] (d) Graphene nanoribbon fabrication SPM based lithography.

**References :**

**1**.	M. Eizenberg and J. M. Blakely, Surf Sci 82 (1), 228 (1979)
**2**.	M. Eizenberg and J. M. Blakely, J Chem Phys 71 (8), 3467 (1979)
**3**.	C. Oshima and A. Nagashima, J Phys-Condens Mat 9 (1), 1 (1997)
**4**.	K. S. Novoselov, A. K. Geim, S. V. Morozov, D. Jiang, Y. Zhang, S. V. Dubonos, I. V. Grigorieva and A. A. Firsov, Science 306 (5296), 666 (2004)
**5**.	C. Lee, X. D. Wei, J. W. Kysar and J. Hone, Science 321 (5887), 385 (2008)
**6**.	A. A. Balandin, S. Ghosh, W. Z. Bao, I. Calizo, D. Teweldebrhan, F. Miao and C. N. Lau, Nano Lett 8 (3), 902 (2008)
**7**.	A. B. Kuzmenko, E. van Heumen, F. Carbone and D. van der Marel, Phys Rev Lett 100 (Copyright (C) 2010 The American Physical Society), 117401 (2008)
**8**.	A. K. Geim and K. S. Novoselov, Nat Mater 6 (3), 183 (2007)
**9**.	M. D. Stoller, S. J. Park, Y. W. Zhu, J. H. An and R. S. Ruoff, Nano Lett 8 (10), 3498 (2008)
**10**.	K. S. Novoselov, Z. Jiang, Y. Zhang, S. V. Morozov, H. L. Stormer, U. Zeitler, J. C. Maan, G. S. Boebinger, P. Kim and A. K. Geim, Science, 1137201 (2007)
**11**.	S. V. Morozov, K. S. Novoselov, M. I. Katsnelson, F. Schedin, D. C. Elias, J. A. Jaszczak and A. K. Geim, Phys Rev Lett 100 (1),  (2008)
**12**.	C. R. Dean, A. F. Young, MericI, LeeC, WangL, SorgenfreiS, WatanabeK, TaniguchiT, KimP, K. L. Shepard and HoneJ, Nat Nano advance online publication,  (2010)
**13**.	S. Stankovich, D. A. Dikin, G. H. B. Dommett, K. M. Kohlhaas, E. J. Zimney, E. A. Stach, R. D. Piner, S. T. Nguyen and R. S. Ruoff, Nature 442 (7100), 282 (2006)
**14**.	W. Hong, H. Bai, Y. Xu, Z. Yao, Z. Gu and G. Shi, The Journal of Physical Chemistry C 114 (4), 1822 (2010)
**15**.	D. Choi, D. Wang, V. V. Viswanathan, I.-T. Bae, W. Wang, Z. Nie, J.-G. Zhang, G. L. Graff, J. Liu, Z. Yang and T. Duong, Electrochemistry Communications 12 (3), 378 (2010)
**16**.	R. Murali, Y. Yang, K. Brenner, T. Beck and J. D. Meindl, Appl Phys Lett 94 (24), 243114 (2009)
**17**.	L. Gomez De Arco, Y. Zhang, C. W. Schlenker, K. Ryu, M. E. Thompson and C. Zhou, ACS Nano,  (2010)
**18**.	X. S. Li, Y. W. Zhu, W. W. Cai, M. Borysiak, B. Y. Han, D. Chen, R. D. Piner, L. Colombo and R. S. Ruoff, Nano Lett 9 (12), 4359 (2009)
**19**.	P. Blake, P. D. Brimicombe, R. R. Nair, T. J. Booth, D. Jiang, F. Schedin, L. A. Ponomarenko, S. V. Morozov, H. F. Gleeson, E. W. Hill, A. K. Geim and K. S. Novoselov, Nano Lett 8 (6), 1704 (2008)
**20**.	X. Wang, L. Zhi and K. Mullen, Nano Lett 8 (1), 323 (2007)
**21**.	G. Eda, G. Fanchini and M. Chhowalla, Nat Nano 3 (5), 270 (2008)
**22**.	X. Lu, M. Yu, H. Huang and R. S. Ruoff, Nanotechnology 10 (3), 269 (1999)
**23**.	Y. Zhang, J. P. Small, W. V. Pontius and P. Kim, Applied Physics Letters 86 (7), 073104 (2005)
**24**.	T. Aizawa, R. Souda, Y. Ishizawa, H. Hirano, T. Yamada, K. I. Tanaka and C. Oshima, Surf Sci 237 (1-3), 194 (1990)
**25**.	Y. Gamo, A. Nagashima, M. Wakabayashi, M. Terai and C. Oshima, Surf Sci 374 (1-3), 61 (1997)
**26**.	D. Farias, K. H. Rieder, A. M. Shikin, V. K. Adamchuk, T. Tanaka and C. Oshima, Surf Sci 454, 437 (2000)
**27**.	A. Nagashima, N. Tejima and C. Oshima, Phys Rev B 50 (23), 17487 (1994)
**28**.	E. Rokuta, Y. Hasegawa, A. Itoh, K. Yamashita, T. Tanaka, S. Otani and C. Oshima, Surf Sci 428, 97 (1999)

29.  Y. Souzu and M. Tsukada, Surface Science 326 (1-2), 42 (1995)
30.  J. C. Hamilton and J. M. Blakely, Surface Science 91 (1), 199 (1980)
31.  L. C. Isett and J. M. Blakely, Surface Science 58 (2), 397 (1976)
32.  H. Zi-Pu, D. F. Ogletree, M. A. Van Hove and G. A. Somorjai, Surface Science 180 (2-3), 433 (1987)
33.  F. J. Himpsel, K. Christmann, P. Heimann, D. E. Eastman and P. J. Feibelman, Surface Science 115 (3), L159 (1982)
34.  A. T. N'Diaye, S. Bleikamp, P. J. Feibelman and T. Michely, Physical Review Letters 97 (Copyright (C) 2010 The American Physical Society), 215501 (2006)
35.  J. Coraux, A. T. N`Diaye, C. Busse and T. Michely, Nano Letters 8 (2), 565 (2008)
36.  H. Ueta, M. Saida, C. Nakai, Y. Yamada, M. Sasaki and S. Yamamoto, Surface Science 560 (1-3), 183 (2004)
37.  D. E. Starr, E. M. Pazhetnov, A. I. Stadnichenko, A. I. Boronin and S. K. Shaikhutdinov, Surface Science 600 (13), 2688 (2006)
38.  S. Marchini, S. Günther and J. Wintterlin, Physical Review B 76 (Copyright (C) 2010 The American Physical Society), 075429 (2007)
39.  N. Gall', E. Rut'kov and A. Tontegode, Physics of the Solid State 46 (2), 371 (2004)
40.  I. Makarenko, A. Titkov, Z. Waqar, P. Dumas, E. Rut'kov and N. Gall', Physics of the Solid State 49 (2), 371 (2007)
41.  A. L. Vázquez de Parga, F. Calleja, B. Borca, M. C. G. Passeggi, J. J. Hinarejos, F. Guinea and R. Miranda, Physical Review Letters 100 (Copyright (C) 2010 The American Physical Society), 056807 (2008)
42.  A. Starodubov, M. Medvetskii, A. Shikin and V. Adamchuk, Physics of the Solid State 46 (7), 1340 (2004)
43.  E. Loginova, N. C. Bartelt, P. J. Feibelman and K. F. McCarty, New Journal of Physics (9), 093026 (2008)
44.  T. A. Land, T. Michely, R. J. Behm, J. C. Hemminger and G. Comsa, Surface Science 264 (3), 261 (1992)
45.  E. Loginova, N. C. Bartelt, P. J. Feibelman and K. F. McCarty, New Journal of Physics (6), 063046 (2009)
46.  K. F. McCarty, P. J. Feibelman, E. Loginova and N. C. Bartelt, Carbon 47 (7), 1806 (2009)
47.  A. Gruneis, K. Kummer and D. V. Vyalikh, New Journal of Physics (7), 073050 (2009)
48.  Y. Pan, H. Zhang, D. Shi, J. Sun, S. Du, F. Liu and H.-j. Gao, Advanced Materials 21 (27), NA (2009)
49.  J. Vaari, J. Lahtinen and P. Hautojärvi, Catalysis Letters 44 (1), 43 (1997)
50.  H. Zhang, Q. Fu, Y. Cui, D. Tan and X. Bao, The Journal of Physical Chemistry C 113 (19), 8296 (2009)
51.  L. J. Ci, L. Song, D. Jariwala, A. L. Elias, W. Gao, M. Terrones and P. M. Ajayan, Adv Mater 21 (44), 4487 (2009)
52.  X. Liang, A. S. P. Chang, Y. Zhang, B. D. Harteneck, H. Choo, D. L. Olynick and S. Cabrini, Nano Letters 9 (1), 467 (2008)
53.  X. Liang, Z. Fu and S. Y. Chou, Nano Letters 7 (12), 3840 (2007)
54.  J.-H. Chen, M. Ishigami, C. Jang, D. R. Hines, M. S. Fuhrer and E. D. Williams, Advanced Materials 19 (21), 3623 (2007)
55.  H. Vincent and et al., Nanotechnology 19 (45), 455601 (2008)
56.  L.-H. Liu and M. Yan, Nano Letters 9 (9), 3375 (2009)
57.  L. Song, L. Ci, W. Gao and P. M. Ajayan, ACS Nano 3 (6), 1353 (2009)
58.  Z. H. Ni, H. M. Wang, J. Kasim, H. M. Fan, T. Yu, Y. H. Wu, Y. P. Feng and Z. X. Shen, Nano Letters 7 (9), 2758 (2007)
59.  A. C. Ferrari, J. C. Meyer, V. Scardaci, C. Casiraghi, M. Lazzeri, F. Mauri, S. Piscanec, D. Jiang, K. S. Novoselov, S. Roth and A. K. Geim, Physical Review Letters 97 (18), 187401 (2006)

60. A. C. Ferrari, Solid State Communications 143 (1-2), 47 (2007)
61. Z. Ni, Y. Wang, T. Yu and Z. Shen, Nano Research 1 (4), 273 (2008)
62. D. A. Dikin, S. Stankovich, E. J. Zimney, R. D. Piner, G. H. B. Dommett, G. Evmenenko, S. T. Nguyen and R. S. Ruoff, Nature 448 (7152), 457 (2007)
63. S. Park, K.-S. Lee, G. Bozoklu, W. Cai, S. T. Nguyen and R. S. Ruoff, ACS Nano 2 (3), 572 (2008)
64. D. Li, M. B. Muller, S. Gilje, R. B. Kaner and G. G. Wallace, Nature Nanotech. 3, 101 (2008)
65. Y. Xu, H. Bai, G. Lu, C. Li and G. Shi, J. Am. Chem. Soc. 130, 5856 (2008)
66. S. Park, J. H. An, R. D. Piner, I. Jung, D. X. Yang, A. Velamakanni, S. T. Nguyen and R. S. Ruoff, Chem Mater 20 (21), 6592 (2008)
67. H. Chen, M. B. Muller, K. J. Gilmore, G. G. Wallace and D. Li, Adv. Mater. 20, 3557 (2008)
68. B. C. Brodie, Ann. Chim. Phys. 59, 466 (1860)
69. W. S. Hummers and R. E. Offeman, Journal of the American Chemical Society 80 (6), 1339 (1958)
70. L. Staudenmaier, Berichte der deutschen chemischen Gesellschaft 31 (2), 1481 (1898)
71. A. Buchsteiner, A. Lerf and J. Pieper, J. Phys. Chem. B 110, 22328 (2006)
72. W. Cai, R. D. Piner, F. J. Stadermann, S. Park, M. A. Shaibat, Y. Ishii, D. Yang, A. Velamakanni, S. J. An, M. Stoller, J. An, D. Chen and R. S. Ruoff, Science 321 (5897), 1815 (2008)
73. W. Gao, L. B. Alemany, L. Ci and P. M. Ajayan, Nat Chem 1 (5), 403 (2009)
74. S. Stankovich, J. Mater. Chem. 16, 155 (2006)
75. I. Jung, Nano Lett. 7, 3569 (2007)
76. S. Park and R. S. Ruoff, Nat Nano 4 (4), 217 (2009)
77. S. Niyogi, E. Bekyarova, M. E. Itkis, J. L. McWilliams, M. A. Hamon and R. C. Haddon, Journal of the American Chemical Society 128 (24), 7720 (2006)
78. K. A. Worsley, Chem. Phys. Lett. 445, 51 (2007)
79. J. R. Lomeda, C. D. Doyle, D. V. Kosynkin, W. F. Hwang and J. M. Tour, J. Am. Chem. Soc. 130, 16201 (2008)
80. R. Muszynski, B. Seger and P. V. Kamat, J. Phys. Chem. C 112, 5263 (2008)
81. G. Williams, B. Serger and P. V. Kamat, ACS Nano 2, 1487 (2008)
82. Y. Geng, Q. B. Zheng and J. K. Kim, J Nanosci Nanotechno 11 (2), 1084 (2011)
83. C. Valles, J. Am. Chem. Soc. 130, 15802 (2008)
84. X. Li, X. Wang, L. Zhang, S. Lee and H. Dai, Science 319, 1229 (2008)
85. X. Li, Nature Nanotech. 3, 538 (2008)
86. J. H. Dong, B. Q. Zeng, Y. C. Lan, S. K. Tian, Y. Shan, X. C. Liu, Z. H. Yang, H. Wang and Z. F. Ren, J Nanosci Nanotechno 10 (8), 5051 (2010)
87. R. Hao, W. Qian, L. Zhang and Y. Hou, Chem. Commun., 6576 (2008)
88. N. Liu, Adv. Funct. Mater. 18, 1518 (2008)
89. Z. Liu, C. W. Fan, L. Chen and A. N. Cao, J Nanosci Nanotechno 10 (11), 7382 (2010)
90. Y. T. Liang and M. C. Hersam, J Am Chem Soc 132 (50), 17661 (2010)
91. A. A. Green and M. C. Hersam, The Journal of Physical Chemistry Letters 1 (2), 544 (2009)
92. A. A. Green and M. C. Hersam, Nano Lett 9 (12), 4031 (2009)
93. Z. Liu, J. T. Robinson, X. Sun and H. Dai, J. Am. Chem. Soc. 130, 10876 (2008)
94. T. Aizawa, Y. Hwang, W. Hayami, R. Souda, S. Otani and Y. Ishizawa, Surface Science 260 (1-3), 311 (1992)
95. T. Aizawa, R. Souda, S. Otani, Y. Ishizawa and C. Oshima, Phys Rev B 42 (18), 11469 (1990)
96. T. Aizawa, R. Souda, S. Otani, Y. Ishizawa and C. Oshima, Phys Rev Lett 64 (7), 768 (1990)
97. Y. Hwang, T. Aizawa, W. Hayami, S. Otani, Y. Ishizawa and S. J. Park, Solid State Communications 81 (5), 397 (1992)
98. H. Itoh, T. Ichinose, C. Oshima, T. Ichinokawa and T. Aizawa, Surf Sci 254 (1-3), L437 (1991)

**99**.	K. Kobayashi and M. Tsukada, Physical Review B 49 (Copyright (C) 2010 The American Physical Society), 7660 (1994)
**100**.	A. Nagashima, K. Nuka, H. Itoh, T. Ichinokawa, C. Oshima and S. Otani, Surface Science 291 (1-2), 93 (1993)
**101**.	A. Nagashima, K. Nuka, K. Satoh, H. Itoh, T. Ichinokawa, C. Oshima and S. Otani, Surf Sci 287, 609 (1993)
**102**.	C. Berger, Z. Song, T. Li, X. Li, A. Y. Ogbazghi, R. Feng, Z. Dai, A. N. Marchenkov, E. H. Conrad, P. N. First and W. A. de Heer, The Journal of Physical Chemistry B 108 (52), 19912 (2004)
**103**.	K. V. Emtsev, A. Bostwick, K. Horn, J. Jobst, G. L. Kellogg, L. Ley, J. L. McChesney, T. Ohta, S. A. Reshanov, J. Rohrl, E. Rotenberg, A. K. Schmid, D. Waldmann, H. B. Weber and T. Seyller, Nat Mater 8 (3), 203 (2009)
**104**.	R. M. Tromp and J. B. Hannon, Physical Review Letters 102 (Copyright (C) 2010 The American Physical Society), 106104 (2009)
**105**.	M. Hupalo, E. H. Conrad and M. C. Tringides, Physical Review B 80 (Copyright (C) 2010 The American Physical Society), 041401 (2009)
**106**.	J. Robinson, X. Weng, K. Trumbull, R. Cavalero, M. Wetherington, E. Frantz, M. LaBella, Z. Hughes, M. Fanton and D. Snyder, ACS Nano 4 (1), 153 (2009)
**107**.	W. Norimatsu and M. Kusunoki, J Nanosci Nanotechno 10 (6), 3884 (2010)
**108**.	E. Moreau, F. J. Ferrer, D. Vignaud, S. Godey and X. Wallart, physica status solidi (a) 207 (2), 300 (2010)
**109**.	A. Al-Temimy, C. Riedl and U. Starke, Applied Physics Letters 95 (23), 231907 (2009)
**110**.	H. Yanagisawa, T. Tanaka, Y. Ishida, M. Matsue, E. Rokuta, S. Otani and C. Oshima, Surface and Interface Analysis 37 (2), 133 (2005)
**111**.	D. Wei, Y. Liu, Y. Wang, H. Zhang, L. Huang and G. Yu, Nano Letters 9 (5), 1752 (2009)
**112**.	X. S. Li, W. W. Cai, J. H. An, S. Kim, J. Nah, D. X. Yang, R. Piner, A. Velamakanni, I. Jung, E. Tutuc, S. K. Banerjee, L. Colombo and R. S. Ruoff, Science 324 (5932), 1312 (2009)
**113**.	A. Reina, X. Jia, J. Ho, D. Nezich, H. Son, V. Bulovic, M. S. Dresselhaus and J. Kong, Nano Letters 9 (1), 30 (2008)
**114**.	K. S. Kim, Y. Zhao, H. Jang, S. Y. Lee, J. M. Kim, K. S. Kim, J.-H. Ahn, P. Kim, J.-Y. Choi and B. H. Hong, Nature 457 (7230), 706 (2009)
**115**.	Q. Yu, J. Lian, S. Siriponglert, H. Li, Y. P. Chen and S.-S. Pei, Applied Physics Letters 93 (11), 113103 (2008)
**116**.	A. N. Obraztsov, E. A. Obraztsova, A. V. Tyurnina and A. A. Zolotukhin, Carbon 45 (10), 2017 (2007)
**117**.	S. J. Chae, F. Güne, scedil, K. K. Kim, E. S. Kim, G. H. Han, S. M. Kim, H.-J. Shin, S.-M. Yoon, J.-Y. Choi, M. H. Park, C. W. Yang, D. Pribat and Y. H. Lee, Advanced Materials 21 (22), 2328 (2009)
**118**.	T. Stefan and et al., Nanotechnology 21 (1), 015601 (2010)
**119**.	A. Reina, S. Thiele, X. Jia, S. Bhaviripudi, M. Dresselhaus, J. Schaefer and J. Kong, Nano Research 2 (6), 509 (2009)
**120**.	M. P. Levendorf, C. S. Ruiz-Vargas, S. Garg and J. Park, Nano Letters 9 (12), 4479 (2009)
**121**.	S. Bhaviripudi, X. Jia, M. S. Dresselhaus and J. Kong, Nano Lett 10 (10), 4128 (2010)
**122**.	X. Li, C. W. Magnuson, A. Venugopal, J. An, J. W. Suk, B. Han, M. Borysiak, W. Cai, A. Velamakanni, Y. Zhu, L. Fu, E. M. Vogel, E. Voelkl, L. Colombo and R. S. Ruoff, Nano Lett 10 (11), 4328 (2010)
**123**.	X. Li, C. W. Magnuson, A. Venugopal, R. M. Tromp, J. B. Hannon, E. M. Vogel, L. Colombo and R. S. Ruoff, J Am Chem Soc 133 (9), 2816 (2011)
**124**.	A. Srivastava, C. Galande, L. Ci, L. Song, C. Rai, D. Jariwala, K. F. Kelly and P. M. Ajayan, Chem Mater, (2010)

**125**. Z. Sun, Z. Yan, J. Yao, E. Beitler, Y. Zhu and J. M. Tour, Nature 468 (7323), 549 (2010)
**126**. Y. Lee, S. Bae, H. Jang, S. Jang, S.-E. Zhu, S. H. Sim, Y. I. Song, B. H. Hong and J.-H. Ahn, Nano Letters 10 (2), 490 (2010)
**127**. K. S. Subrahmanyam, L. S. Panchakarla, A. Govindaraj and C. N. R. Rao, The Journal of Physical Chemistry C 113 (11), 4257 (2009)
**128**. W. Zhiyong and et al., Nanotechnology 21 (17), 175602 (2010)
**129**. L. S. Panchakarla, K. S. Subrahmanyam, S. K. Saha, A. Govindaraj, H. R. Krishnamurthy, U. V. Waghmare and C. N. R. Rao, Advanced Materials 21 (46), 4726 (2009)
**130**. A. Dato, V. Radmilovic, Z. Lee, J. Phillips and M. Frenklach, Nano Letters 8 (7), 2012 (2008)
**131**. E. Dervishi, Z. Li, F. Watanabe, A. Biswas, Y. Xu, A. R. Biris, V. Saini and A. S. Biris, Chemical Communications (27), 4061 (2009)
**132**. J. Campos-Delgado, J. M. Romo-Herrera, X. Jia, D. A. Cullen, H. Muramatsu, Y. A. Kim, T. Hayashi, Z. Ren, D. J. Smith, Y. Okuno, T. Ohba, H. Kanoh, K. Kaneko, M. Endo, H. Terrones, M. S. Dresselhaus and M. Terrones, Nano Letters 8 (9), 2773 (2008)
**133**. M. Choucair, P. Thordarson and J. A. Stride, Nat Nano 4 (1), 30 (2009)
**134**. C. N. R. Rao, A. K. Sood, R. Voggu and K. S. Subrahmanyam, The Journal of Physical Chemistry Letters 1 (2), 572 (2010)
**135**. C. Rao, A. Sood, K. Subrahmanyam and A. Govindaraj, Angewandte Chemie International Edition 48 (42), 7752 (2009)
**136**. M. J. Allen, V. C. Tung and R. B. Kaner, Chem Rev 110 (1), 132 (2009)
**137**. T. Kawasaki, Surf. Rev. Lett. 9, 1459 (2002)
**138**. G. Giovannetti, P. A. Khomyakov, G. Brocks, P. J. Kelly and J. van den Brink, Physical Review B 76 (Copyright (C) 2010 The American Physical Society), 073103 (2007)
**139**. P. Shemella and S. K. Nayak, Applied Physics Letters 94 (3), 032101 (2009)
**140**. S. Y. Zhou, G. H. Gweon, A. V. Fedorov, P. N. First, W. A. de Heer, D. H. Lee, F. Guinea, A. H. Castro Neto and A. Lanzara, Nat Mater 6 (10), 770 (2007)
**141**. S. S. Yu, W. T. Zheng, Q. B. Wen and Q. Jiang, Carbon 46 (3), 537 (2008)
**142**. Y. Li, Z. Zhou, P. Shen and Z. Chen, ACS Nano 3 (7), 1952 (2009)
**143**. A. Lherbier, X. Blase, Y.-M. Niquet, F. Triozon and S. Roche, Physical Review Letters 101 (Copyright (C) 2010 The American Physical Society), 036808 (2008)
**144**. X. H. Zheng, X. L. Wang, T. A. Abtew and Z. Zeng, The Journal of Physical Chemistry C 114 (9), 4190 (2010)
**145**. R. Peköz and S. Erkoç, Physica E: Low-dimensional Systems and Nanostructures 42 (2), 110 (2009)
**146**. Y. Shan Sheng, Z. Wei Tao and J. Qing, Nanotechnology, IEEE Transactions on 9 (1), 78 (2010)
**147**. F. Cervantes-Sodi, G. Csányi, S. Piscanec and A. C. Ferrari, Physical Review B 77 (Copyright (C) 2010 The American Physical Society), 165427 (2008)
**148**. N. Li, Z. Wang, K. Zhao, Z. Shi, Z. Gu and S. Xu, Carbon 48 (1), 255 (2010)
**149**. X. Ma, Q. Wang, L. Q. Chen, W. Cermignani, H. H. Schobert and C. G. Pantano, Carbon 35 (10-11), 1517 (1997)
**150**. S. Dutta and S. K. Pati, The Journal of Physical Chemistry B 112 (5), 1333 (2008)
**151**. T. B. Martins, R. H. Miwa, A. J. R. da Silva and A. Fazzio, Physical Review Letters 98 (Copyright (C) 2010 The American Physical Society), 196803 (2007)
**152**. A. Quandt, C. Özdo, gbreve, an, J. Kunstmann and H. Fehske, physica status solidi (b) 245 (10), 2077 (2008)
**153**. L. Ci, L. Song, C. Jin, D. Jariwala, D. Wu, Y. Li, A. Srivastava, Z. F. Wang, K. Storr, L. Balicas, F. Liu and P. M. Ajayan, Nat Mater 9 (5), 430 (2010)
**154**. J. Li and V. B. Shenoy, Appl Phys Lett 98 (1), 013105 (2011)

155. M. Endo, T. Hayashi, S.-H. Hong, T. Enoki and M. S. Dresselhaus, Journal of Applied Physics 90 (11), 5670 (2001)
156. A. Y. Liu, R. M. Wentzcovitch and M. L. Cohen, Phys. Rev. B 39, 1760 (1989)
157. Y. Miyamoto, A. Rubio, M. L. Cohen and S. G. Louie, Phys. Rev. B 50, 4976 (1994)
158. E. Hernandez, C. Goze, P. Bernier and A. Rubio, Phys. Rev. Lett. 80, 4502 (1998)
159. Z. Weng-Sieh, K. Cherrey, N. G. Chopra, X. Blase, Y. Miyamoto, A. Rubio, M. L. Cohen, S. G. Louie, A. Zettl and R. Gronsky, Physical Review B 51 (Copyright (C) 2010 The American Physical Society), 11229 (1995)
160. D. Golberg, P. Dorozhkin, Y. Bando and Z. C. Dong, MRS Bull. 29, 38 (2004)
161. R. B. Kaner, J. Kouvetakis, C. E. Warble, M. L. Sattler and N. Bartlett, Mater. Res. Bull. 22, 399 (1987)
162. M. Kawaguchi, T. Kawashima and T. Nakajima, Chem. Mater. 8, 1197 (1996)
163. J. Yu, E. G. Wang, J. Ahn, S. F. Yoon, Q. Zhang, J. Cui and M. B. Yu, Journal of Applied Physics 87 (8), 4022 (2000)
164. K. Yuge, Phys. Rev. B 79, 144109 (2009)
165. S. Enouz, O. Stéphan, J.-L. Cochon, C. Colliex and A. Loiseau, Nano Letters 7 (7), 1856 (2007)
166. V. V. Ivanovskaya, A. Zobelli, O. Stéphan, P. R. Briddon and C. Colliex, The Journal of Physical Chemistry C 113 (38), 16603 (2009)
167. Y. Ding, Y. Wang and J. Ni, Applied Physics Letters 95 (12), 123105 (2009)
168. K. Nakada, M. Fujita, G. Dresselhaus and M. S. Dresselhaus, Phys Rev B 54 (24), 17954 (1996)
169. K. Wakabayashi, M. Fujita, H. Ajiki and M. Sigrist, Phys Rev B 59 (12), 8271 (1999)
170. M. Fujita, K. Wakabayashi, K. Nakada and K. Kusakabe, J. Phys. Soc. Jpn. 65, 1920 (1996)
171. X. L. Li, Science 319, 1229 (2008)
172. C. Berger, Z. Song, X. Li, X. Wu, N. Brown, C. Naud, D. Mayou, T. Li, J. Hass, A. N. Marchenkov, E. H. Conrad, P. N. First and W. A. de Heer, Science 312 (5777), 1191 (2006)
173. L. A. Ponomarenko, F. Schedin, M. I. Katsnelson, R. Yang, E. W. Hill, K. S. Novoselov and A. K. Geim, Science 320 (5874), 356 (2008)
174. W. Liu, Z. F. Wang, Q. W. Shi, J. Yang and F. Liu, Phys. Rev. B 80, 233405 (2009)
175. T. G. Pedersen, C. Flindt, J. Pedersen, N. A. Mortensen, A.-P. Jauho and K. Pedersen, Physical Review Letters 100 (Copyright (C) 2010 The American Physical Society), 136804 (2008)
176. V. Nenad and et al., Physical Review B 81 (4), 041408 (2010)
177. J. A. Fürst and et al., New Journal of Physics 11 (9), 095020 (2009)
178. J. Eroms and D. Weiss, New Journal of Physics 11 (9), 095021 (2009)
179. J. Bai, X. Zhong, S. Jiang, Y. Huang and X. Duan, Nat Nano 5 (3), 190 (2010)
180. M. Kim, N. S. Safron, E. Han, M. S. Arnold and P. Gopalan, Nano Lett 10 (4), 1125 (2010)
181. R. Balog, B. Jorgensen, L. Nilsson, M. Andersen, E. Rienks, M. Bianchi, M. Fanetti, E. Laegsgaard, A. Baraldi, S. Lizzit, Z. Sljivancanin, F. Besenbacher, B. Hammer, T. G. Pedersen, P. Hofmann and L. Hornekaer, Nat Mater advance online publication, (2010)
182. J. Kotakoski, A. V. Krasheninnikov, U. Kaiser and J. C. Meyer, Phys Rev Lett 106 (10), 105505 (2011)
183. T. O. Wehling, K. S. Novoselov, S. V. Morozov, E. E. Vdovin, M. I. Katsnelson, A. K. Geim and A. I. Lichtenstein, Nano Lett 8 (1), 173 (2008)
184. D. C. Elias, R. R. Nair, T. M. G. Mohiuddin, S. V. Morozov, P. Blake, M. P. Halsall, A. C. Ferrari, D. W. Boukhvalov, M. I. Katsnelson, A. K. Geim and K. S. Novoselov, Science 323 (5914), 610 (2009)
185. Hern, aacute, A. D. ndez-Nieves, B. Partoens and F. M. Peeters, Phys Rev B 82 (16), 165412 (2010)
186. J.-H. Lee and J. C. Grossman, Appl Phys Lett 97 (13), 133102 (2010)
187. M. Yang, A. Nurbawono, C. Zhang, Y. P. Feng and Ariando, Appl Phys Lett 96 (19), 193115 (2010)

188. R. R. Nair, W. Ren, R. Jalil, I. Riaz, V. G. Kravets, L. Britnell, P. Blake, F. Schedin, A. S. Mayorov, S. Yuan, M. I. Katsnelson, H.-M. Cheng, W. Strupinski, L. G. Bulusheva, A. V. Okotrub, I. V. Grigorieva, A. N. Grigorenko, K. S. Novoselov and A. K. Geim, Small 6 (24), 2877 (2010)
189. J. T. Robinson, J. S. Burgess, C. E. Junkermeier, S. C. Badescu, T. L. Reinecke, F. K. Perkins, M. K. Zalalutdniov, J. W. Baldwin, J. C. Culbertson, P. E. Sheehan and E. S. Snow, Nano Lett 10 (8), 3001 (2010)
190. M. Y. Han, B. Ozylmaz, Y. Zhang and P. Kim, Phys. Rev. Lett. 98, 206805 (2007)
191. Z. Chen, Y.-M. Lin, M. J. Rooks and P. Avouris, Physica E: Low-dimensional Systems and Nanostructures 40 (2), 228 (2007)
192. J. Bai, X. Duan and Y. Huang, Nano Lett 9 (5), 2083 (2009)
193. L. Liao, J. Bai, R. Cheng, Y.-C. Lin, S. Jiang, Y. Huang and X. Duan, Nano Lett 10 (5), 1917 (2010)
194. D. C. Bell and et al., Nanotechnology 20 (45), 455301 (2009)
195. A. Dimiev, D. V. Kosynkin, A. Sinitskii, A. Slesarev, Z. Sun and J. M. Tour, Science 331 (6021), 1168 (2011)
196. L. J. Ci, Nano Res. 1, 116 (2008)
197. S. S. Datta, D. R. Strachan, S. M. Khamis and A. T. C. Johnson, Nano Lett. 8, 1912 (2008)
198. N. Severin, S. Kirstein, I. M. Sokolov and J. P. Rabe, Nano Lett 9 (1), 457 (2008)
199. F. Schäffel, J. Warner, A. Bachmatiuk, B. Rellinghaus, B. Büchner, L. Schultz and M. Rümmeli, Nano Res 2 (9), 695 (2009)
200. F. Schäffel, J. H. Warner, A. Bachmatiuk, B. Rellinghaus, B. Büchner, L. Schultz and M. H. Rümmeli, physica status solidi (b) 246 (11-12), 2540 (2009)
201. J. H. Warner, M. H. Rümmeli, A. Bachmatiuk, M. Wilson and B. Büchner, ACS Nano 4 (1), 470 (2009)
202. A. Tomita and Y. Tamai, The Journal of Physical Chemistry 78 (22), 2254 (1974)
203. C. W. Keep, S. Terry and M. Wells, Journal of Catalysis 66 (2), 451 (1980)
204. R. T. K. Baker, R. D. Sherwood and E. G. Derouane, Journal of Catalysis 75 (2), 382 (1982)
205. P. J. Goethel and R. T. Yang, Journal of Catalysis 108 (2), 356 (1987)
206. L. C. Campos, V. R. Manfrinato, J. D. Sanchez-Yamagishi, J. Kong and P. Jarillo-Herrero, Nano Lett. 9, 2600 (2009)
207. M. Ringger, H. R. Hidber, R. Schlogl, P. Oelhafen and H. J. Guntherodt, Appl Phys Lett 46 (9), 832 (1985)
208. R. L. McCarley, S. A. Hendricks and A. J. Bard, The Journal of Physical Chemistry 96 (25), 10089 (1992)
209. L. Tapaszto, G. Dobrik, P. Lambin and L. P. Biro, Nat Nano 3 (7), 397 (2008)
210. G. Dobrik, L. Tapasztó, P. Nemes-Incze, P. Lambin and L. P. Biró, physica status solidi (b) 247 (4), 896 (2010)
211. L. Weng, L. Zhang, Y. P. Chen and L. P. Rokhinson, Appl Phys Lett 93 (9), 093107 (2008)
212. A. J. M. Giesbers, U. Zeitler, S. Neubeck, F. Freitag, K. S. Novoselov and J. C. Maan, Solid State Commun 147 (9-10), 366 (2008)
213. A. Knoll, P. Bächtold, J. Bonan, G. Cherubini, M. Despont, U. Drechsler, U. Dürig, B. Gotsmann, W. Häberle, C. Hagleitner, D. Jubin, M. A. Lantz, A. Pantazi, H. Pozidis, H. Rothuizen, A. Sebastian, R. Stutz, P. Vettiger, D. Wiesmann and E. S. Eleftheriou, Microelectronic Engineering 83 (4-9), 1692
214. J. Haaheim and O. A. Nafday, Scanning 30 (2), 137 (2008)
215. L. P. Biró and P. Lambin, Carbon In Press, Corrected Proof,
216. L. Y. Jiao, L. Zhang, X. R. Wang, G. Diankov and H. J. Dai, Nature 458, 877 (2009)
217. D. V. Kosynkin, A. L. Higginbotham, A. Sinitskii, J. R. Lomeda, A. Dimiev, B. K. Price and J. M. Tour, Nature 458 (7240), 872 (2009)

**218**.	A. L. Elías, A. s. R. Botello-Méndez, D. Meneses-Rodríguez, V. Jehová González, D. Ramírez-González, L. Ci, E. Muñoz-Sandoval, P. M. Ajayan, H. Terrones and M. Terrones, Nano Lett 10 (2), 366 (2009)
**219**.	K. Kim, A. Sussman and A. Zettl, ACS Nano 4 (3), 1362 (2010)
**220**.	W. S. Kim, S. Y. Moon, S. Y. Bang, B. G. Choi, H. Ham, T. Sekino and K. B. Shim, Appl Phys Lett 95 (8), 083103 (2009)
**221**.	A. G. Cano-Márquez, F. J. Rodríguez-Macías, J. Campos-Delgado, C. G. Espinosa-González, F. Tristán-López, D. Ramírez-González, D. A. Cullen, D. J. Smith, M. Terrones and Y. I. Vega-Cantú, Nano Lett 9 (4), 1527 (2009)
**222**.	M. Terrones, ACS Nano 4 (4), 1775 (2010)
**223**.	M. C. a. o. Paiva, W. Xu, M. Fernanda Proença, R. M. Novais, E. Lægsgaard and F. Besenbacher, Nano Lett 10 (5), 1764 (2010)
**224**.	L. Jiao, X. Wang, G. Diankov, H. Wang and H. Dai, Nat Nano 5 (5), 321 (2010)
**225**.	J. Cai, P. Ruffieux, R. Jaafar, M. Bieri, T. Braun, S. Blankenburg, M. Muoth, A. P. Seitsonen, M. Saleh, X. Feng, K. Mullen and R. Fasel, Nature 466 (7305), 470 (2010)
**226**.	SprinkleM, RuanM, HuY, HankinsonJ, M. Rubio Roy, ZhangB, WuX, BergerC and W. A. de Heer, Nat Nano 5 (10), 727 (2010)
**227**.	S. Bae, H. Kim, Y. Lee, X. Xu, J.-S. Park, Y. Zheng, J. Balakrishnan, T. Lei, H. Ri Kim, Y. I. Song, Y.-J. Kim, K. S. Kim, B. Ozyilmaz, J.-H. Ahn, B. H. Hong and S. Iijima, Nat Nano 5 (8), 574 (2010)